\documentclass{article}
\usepackage[utf8]{inputenc}
\usepackage{geometry}
\usepackage{amsmath}
\usepackage{titling}
\usepackage{hyperref}
\usepackage{csquotes}
\usepackage[sorting=none, style=nature]{biblatex}
\usepackage{algpseudocode}
\usepackage{algorithm}
\usepackage{graphicx}
\usepackage{newtxtext,newtxmath}

\begin{document}

\title{Foundations of the WKB Approximation for Models of Cochlear Mechanics in 1- and 2-D}
\author{Brian L. Frost}
%\affiliation{Department of Electrical Engineering, Columbia University, 500 W. 120th St., Mudd 1310, New York, NY 10027, USA.}
%\email{b.frost@columbia.edu}

%\preprint{Frost, JASA}

\date{January, 2024} 

\maketitle

\begin{abstract}
    The Wentzel-Kramers-Brillouin (WKB) approximation is frequently used to explore the mechanics of the cochlea. As opposed to numerical strategies, the WKB approximation facilitates analysis of model results through interpretable closed-form equations, and can be implemented with relative ease. As a result, it has maintained relevance in the study of cochlear mechanics for half of a century. Over this time, it has been used to study a variety of phenomena including the limits of frequency tuning, active displacement amplification within the organ of Corti, feedforward mechanisms in the cochlea, and otoacoustic emissions. Despite this ubiquity, it is challenging to find rigorous exposition of the WKB approximation's formulation, derivation and implementation in cochlear mechanics literature. In this tutorial, I discuss the foundations of the WKB approximation in application to models of cochlear macromechanics in 1-D and 2-D. This  includes mathematical background, rigorous derivation and details of its implementation in software.  
\end{abstract}

\section{Introduction}

Models of 1-D and 2-D macromechanics have offered some of the most significant interpretations of cochlear physiology, both historic and modern. It is intuitive that 3-D models should offer more physically realistic results than 2-D or 1-D models, but this alone implies a ``more the merrier" view of model dimensionality that coincides with quantitative accuracy, but not with frequency of implementation or impact on the field of cochlear mechanics.

{Important results of 1-D models include} the existence and character of stapes-driven traveling waves and the presence of a region of negative damping  \cite{Wegel_Lane_1924,Zwislocki_1980,Peterson_Bogert_1950,Zweig_compromise,Zweig_1991,catastrophe,Elliott_Ku_Lineton_2007} as well as intra-cochlear reflections and otoacoustic emissions (OAEs) \cite{Viergever_1986,de_Boer_Kaernbach_Konig_Schillen_1986,basis,Zweig_1991,Talmadge_Tubis_Long_Piskorski_1998,Talmadge_Tubis_Long_Tong_2000,Shera_Tubis_Talmadge_2005,oae_moleti,oae_sisto}. Qualitative similarity across frequency/space and quantitative similarity in the long-wave region to \textit{in vivo} cochlear responses make 1-D models attractive for the exploration of fundamental macromechanical phenomena. Both implementation and modification of the dynamics to account for features such as nonlinearity and roughness are also far simpler in 1-D models than in 2-D or 3-D models\cite{basis,Talmadge_Tubis_Long_Tong_2000,Talmadge_Tubis_Long_Piskorski_1998,Shera_Tubis_Talmadge_2005}.

On the other hand, 2-D models allow for more physical results in the short-wave and cutoff regions of the cochlear response than 1-D models \cite{Siebert_1973,Allen_1977,Allen_Sondhi_1979,neely,steele_rootfinding,viergever_Book}, allowing more complete exploration of potential {mode-coupling phenomena} in the traveling wave \cite{watts,Watts_2000,Elliott_Ni_Mace_Lineton_2013}. Moreover, 2-D models are able to capture fluid mechanical properties in the scalae that allow for interpretation of energy flow \cite{lighthill_long,lighthill_short,steele_lagrange}, or the manner by which pressure across the scalae is focused at the organ of Corti complex (OCC) to supply energy to the traveling wave \cite{duifhuis_moh,sisto_2021,sisto_2023,Shera_Altoe_2023}, whereas 1-D models describe only the average pressure across the scalae\cite{Olson_2001}\footnote{This concept has been of particular interest in cochlear mechanics since  Olson showed that pressure varies rapidly in space within the scala, and is tuned near the OCC.}.

The macromechanics of the cochlea are generally modeled as boundary value problems (BVPs) where the model equations involve partial differential equations (PDEs) without analytically known solutions, such as the Navier-Stokes equation. Such model equations can be tackled using numerics (e.g. the finite element method), or by making sufficient simplifying assumptions so that approximate analytic solutions can be found -- the scala walls are rigid, the fluids are incompressible, etc. 

Techniques based on the Wentzel-Kramers-Brillouin (WKB) approximation, also known as the Liouville-Green (LG) approximation \cite{Liouville_1837,green,mathews_wkb,Dingle_1975,Winitzki_2005}, {were introduced to the field of cochlear mechanics by Zweig and} are among the most popular for achieving approximate, explicit closed-form solutions for OCC motion, fluid pressure and fluid velocity in a variety of cochlear models that match numerical solutions well across broad frequency and spatial ranges \cite{Zweig_compromise,Zweig_1991,duifhuis_moh,viergever_Book,de_Boer_Kaernbach_Konig_Schillen_1986,deBoer_PhysicsReports,deBoer_Energy,sisto_2021,altoe_2022,sisto_2023,Shera_Altoe_2023}. Closed-form explicit solutions allow for more easily interpreted model results. While exact solutions have been derived and studied for some cochlear models -- e.g. implicit Green's function solutions \cite{Allen_1977,Allen_Sondhi_1979,Mammano_Nobili_1993} or explicit Fourier transform solutions for box models \cite{deboer_block} -- they are not as simple to qualitatively analyze. 

As a nonexhaustive list, 1-D and 2-D WKB approximate solutions have offered: interpretations of limits on cochlear tuning \cite{Zweig_compromise}; interpretations of intracochlear reflections and OAEs \cite{basis,Talmadge_Tubis_Long_Piskorski_1998,Talmadge_Tubis_Long_Tong_2000,Shera_Tubis_Talmadge_2005,oae_sisto,Sisto_Moleti_Shera_2007,Shera_Altoe_2023}; interpretations of traveling wave mode {coupling} \cite{watts,Watts_2000,Elliott_Ni_Mace_Lineton_2013}; interpretations of the impact of active power generation in the cochlea \cite{sisto_2021,sisto_2023}. {These approximations can also be applied to accelerate computations in more complex 3-D cochlear models (e.g. \cite{Yoon2011})}. Robustness, ease of implementation, interpretability and versatility have earned the WKB approximation its persistence in macromechanics modeling over half of a century.

With the passage of time, foundations of WKB techniques have largely disappeared from cochlear mechanics literature; as with any historical method, derivations, assumptions, implementation details and the implications thereof have become implicit. This efficiency is useful for experienced readers, but creates confusion for newer entrants to the field. In the case of WKB, not only are these objects often missing in contemporary literature, but challenging to find in historic literature as well.

The relevance of the WKB approximation in cochlear models, both historical and contemporary, owes it a foundational exposition. Fundamental understanding of the approximation can open the door to adaptations for probing particular questions, with knowledge of its strengths and limitations. As such questions continue to arise with the publication of new data, especially with the advent of optical coherence tomography, this is all the more relevant.

Lastly, the recent passing of Egbert de Boer, Hendrikus Duifhuis and Charles Steele -- three pioneers of the application of the WKB approximation to cochlear mechanics -- suggests a timeliness of such a presentation.

The essence of this tutorial is to present the fundamentals of WKB techniques in linear 1-D and 2-D cochlear mechanics models from an analytic perspective, covering derivations and details of implementation and performance. 

I begin by describing general mathematical details of the WKB approximation agnostic to cochlear applications. This is followed by a description of the 1-D and 2-D BVPs for the box model of the cochlea. Derivations of the 1-D and 2-D WKB solutions to these BVPs follow. I then discuss the theory of the WKB traveling wave subspace (in terms of ``WKB basis functions") most often used in the study of intracochlear reflections and OAEs.

For readers interested in implementation rather than theory, Sec \ref{sec:implement} discusses practical concerns. This includes discussion of several common methods for solving the dispersion relation for 2-D box models, along with details of their performance across frequency and spatial ranges. This is followed by a comparison across methods and to numerical solutions.

\pagebreak
\clearpage
\section{The WKB Approximation}

In this section, I will present the mathematical underpinnings of the WKB approximation. These abstract concepts will be applied to cochlear mechanics models specifically in the following sections. 

Consider a homogeneous linear $n^{\text{th}}$ order ordinary differential equation (ODE) of the form
\begin{equation}
    \epsilon\frac{d^n y}{dx} + a_{n-1}(x)\frac{d^{n-1} y}{dx^{n-1}} + \ldots + a_1(x)\frac{dy}{dx} + a_0(x)y = 0,
    \label{WKBODE}
\end{equation}
where the coefficient functions $a_i$, $i=1,2,\ldots,n-1$ are $n$-times continuously differentiable functions on some interval $I\subset \mathbb{R}$, and $\epsilon\in\mathbb{R}$ is presumed to be small relative to the magnitudes of the other coefficient functions. The coefficient functions may be complex-valued.

Consider an ansatz for a solution to Eqn \ref{WKBODE} as the exponential of a formal power series in $\delta\in\mathbb{R}$,
\begin{equation}
    y(x) = \exp{\bigg[\frac{1}{\delta}\sum_{m=0}^{\infty} \delta^{m}C_m(x)}\bigg],
    \label{WKB_withdelta}
\end{equation}
where $C_m$, $m=0,1,2,\ldots$ are $n$-times continuously differentiable functions on $I$ and it is assumed that the series can be differentiated term-wise \cite{Liouville_1837,green,Robnik_Romanovski_2000,Dingle_1975,Winitzki_2005}. So that the series converges\footnote{In reality, the series often \textit{diverges} and terms will begin to increase after a certain order of $m$. This limits the precision of the approximation, but will not come into play in this tutorial as I will never exceed $m=1$. Detailed discussion is presented in \textit{Winitzky, 2005} \cite{Winitzki_2005}.}, $\delta$ should be small and $C_m$ and their derivatives must fall off exponentially in magnitude across the real line with increasing $m$. That is,
\begin{equation}
    \big|\delta^{m}C_{m+1}(x)\big| \ll \big|\delta^{m-1}C_m(x)\big|,\;\;m=0,1,2,\ldots.
\end{equation}

The ansatz, when plugged into Eqn \ref{WKBODE}, yields a differential equation containing infinitely many unknowns, $C_m$. The \textit{$M^{\text{th}}$-order WKB approximation} is made by truncating this series up to the $M^{\text{th}}$ term. This is valid so long as all terms at indices higher than $M$ are much smaller than 1 on $I$. That is,
\begin{equation}
    \big|\delta^M C_{M+1}(x)\big| \ll 1.
\end{equation}
It is common that the first-order WKB approximation is simply called ``the WKB approximate solution" \cite{mathews_wkb}. 
\vfill
\pagebreak
\clearpage
\section{Cochlear Model}
\label{sec:3d}

\begin{figure}[ht]
    \centering
    \includegraphics[width=\textwidth]{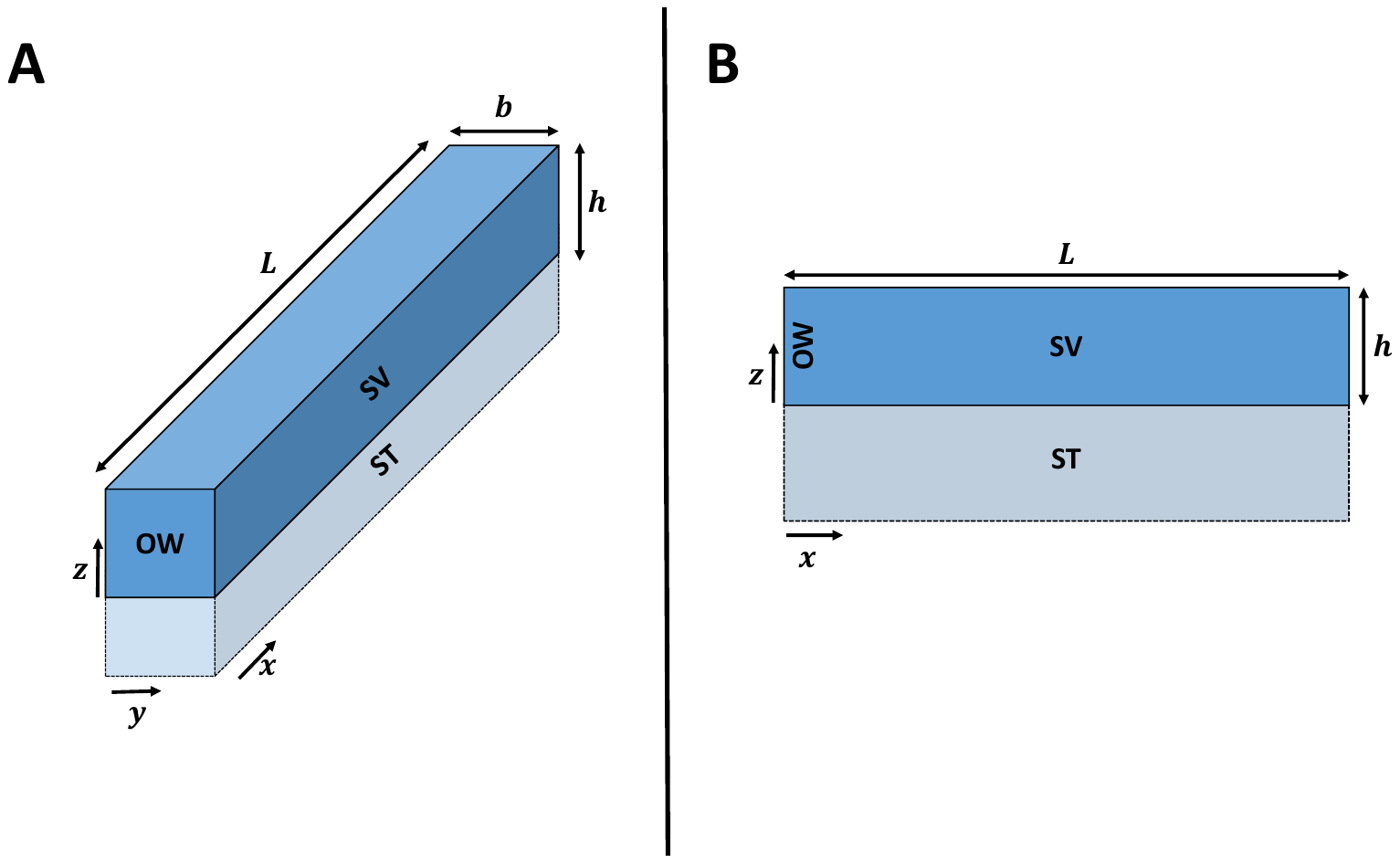}
    \caption{\textbf{A} -- Geometry of the 3-D box model, having length $L$, scala height $h$ and width $b$. The longitudinal, radial and transverse directions are $x$, $y$ and $z$, respectively. \textbf{B} -- Geometry of the corresponding 2-D box model. In the tapered box model, $h$ may be a function of $x$. SV = Scala Vestibuli; ST = Scala Tympani; OW = Oval Window.}
    \label{fig:box}
\end{figure}

The WKB approximation can be applied to any cochlear model described by linear differential equations. In this tutorial, the focus is one popular class of models -- box and tapered box models -- with geometry as shown in Fig \ref{fig:box}. The model is derived by considering the cochlea as uncoiled and containing two scalae -- scala vestibuli (SV) and scala tympani (ST) \cite{bekesy}\footnote{{Reissner's membrane, separating SV and scala media is assumed to be mechanically invisible, so we need not distinguish between all three chambers.}}. The scalae are separated by an infinitesimally thin plate with the flexible OCC being the only portion of this plate capable of movement.

The cochlea's longitudinal axis ($x$) points towards the apex, terminating at the stapes at $x=0$ and the helicotrema at $x=L$. The transverse axis ($z$) points towards SV with the OCC lying at $z=0$. The cross-sectional area of SV and ST are equal to one another\footnote{This assumption can be relaxed and assymetric scalae can be assumed by finding some effective area. Results are largely unchanged.} and vary along the longitudinal axis as $S(x)$. The OCC width varies along the longitudinal axis as $b(x)$. This model simplifies to the common box model when the scala walls are not curved and $S(x)$ and $b(x)$ are constant.

The model can be flattened to 2-D, as is represented geometrically in Fig \ref{fig:box}. This flattening amounts to representation of each quantity as its average over the radial dimension. It appears as a tapered box with height $h(x) = S(x)/b(x)$. Further flattening of the model to 1-D involves representation of all quantities as being only dependent on $x$. This amounts to averaging quantities over transverse space.

\subsection{Boundary Conditions and Assumptions}
\label{sec:conditions}

The boundary conditions are determined based on the following assumptions: 1) fluid does not flow in the normal direction towards or out of the scalae at the walls $z=\pm h$, and at the helicotrema $x=L$, 2) the average pressure at  $x=0$ is some known constant $P_{OW}$, and 3) the OCC is mechanically described by a longitudinally-varying point impedance $Z_{OC}(x)$ (or reciprocally as a point admittance $Y_{OC} = 1/Z_{OC}$)\cite{ViergeverImpedance}. This quantity is complex and frequency-dependent.

It should be noted that the modeled pressure and velocity will vary sinusoidally. Assuming linearity of the model\cite{Kanis_deBoer_1993,Kanis_deBoer_1996,Talmadge_Tubis_Long_Piskorski_1998,altoe_2022}\footnote{The focus of this tutorial is on linear models, but nonlinearity can be implemented in a number of ways (e.g. quasilinear method, formulation of a nonhomogeneous cochlear model, etc.).}, inputs at a given radian frequency $\omega$ will yield model responses at the same frequency. That is, all quantities will be of the form $C(x,z,\omega)e^{j\omega  t}$.  The time-dependence is identical across all quantities and so it will generally be left implicit. 

The fluid pressure is denoted $P(x,z)$, the longitudinal fluid velocity is denoted $\dot{u}(x,z)$ and the transverse fluid velocity is denoted $\dot{w}(x,z)$. The impedance describes the relationship between transmembrane pressure and transverse displacement at $z=0$. Due to the symmetry of the model, transmembrane pressure $p=2P$. 

In the 2-D tapered box model, the boundary conditions can be written as
\begin{align}
    &\frac{1}{h(0)}\int_{0}^{h(0)} P(0,z)\;dz = P_{OW}, \label{owBV}\;\;\\
    &\frac{\partial P}{\partial x}(L,z) = 0, \label{heliBV}\\
    &\frac{\partial P}{\partial z}(x,h(x)) = 0, \label{wallBV}\\
    &P(x,0)  = -\frac{Z_{OC}(x)}{2} \dot{w}(x,0), \label{bvImpedance}
\end{align}
where the negative sign in the impedance relation comes from positive pressure in SV applying a $-z$-direction (downward) force on the OCC. These boundary conditions are equally valid in the 1-D model simply by ignoring dependence on $z$.

An assumption is also made regarding the character of the traveling wave solutions. {With the input pressure appearing at the stapes, the traveling wave will primarily travel towards the apex. In this text, these will be referred to as \textit{apical-traveling waves}. At the helicotrema, reflection will occur and a wave traveling towards the stapes will be generated. In this text, such waves will be referred to as \textit{basal-traveling waves}.}

For the most part, I will assume that {basal-traveling waves are far smaller than apical-traveling waves}. In some models, this is achieved by letting $L\rightarrow \infty$, in which case the WKB solutions will be identical to those arrived at in this tutorial. Basal-traveling waves will be considered in the context of WKB basis functions (Sec \ref{sec:proj}), and this assumption will be removed.

\subsection{Model Equations}

With the geometry described, the model equations can now be developed. The fluid in the scalae is modeled as \textit{incompressible}, \textit{irrotational}, \textit{linear} and \textit{inviscid}. The fluid velocity vector  $\mathbf{v} = (\dot{u}\; \dot{w})^T$ in a volume satisfies the continuity equation, given by
\begin{equation}
    \frac{\partial\rho}{\partial t} + \nabla\cdot(\rho\mathbf{v})= 0,
\end{equation}
where $\rho$ is the fluid density. This represents that within a differential volume, the change in fluid mass in the region is accompanied by an equal and opposite divergence of that fluid into/out of the region.

In an incompressible fluid, the mass of the fluid (and thereby $\rho$) in any region is constant, simplifying the equation to
\begin{equation}
\nabla\cdot\mathbf{v} = 0.
\label{cont}
\end{equation}
An irrotational field is also a conservative field, so the velocity field can be written as the gradient of some scalar field $\phi$. This \textit{velocity potential} is thereby defined by

\begin{equation}
    \nabla\phi = \mathbf{v}.
\end{equation}

Taking the divergence of both sides and applying Eqn \ref{cont} yields the Laplace equation in velocity potential:
\begin{equation}
    \nabla^2\phi = 0.
\end{equation}
The Navier-Stokes equation in an inviscid, incompressible, linear and irrotational fluid is
\begin{equation}
\rho\frac{\partial \mathbf{v}}{\partial t} + \nabla P =\rho\frac{\partial \nabla \phi}{\partial t} + \nabla P=0.
\label{navier}
\end{equation}
This gives 
\begin{equation}
    P = -\rho \dot{\phi},
    \label{ptophi}
\end{equation}
where the overhead dot indicates a partial derivative in time. Taking the Laplacian of both sides and recalling that $\phi$ satisfies the Laplace equation, we arrive at a Laplace equation in $P$:
\begin{equation}
    \nabla^2 P=0.
\end{equation}
The Laplace equation then also holds for transmembrane pressure $p=2P$. 

Another model equation can be derived for the \textit{average} pressure in a cross-section, i.e. for the \textit{1-D model} where pressure and velocity depend only on $x$. Over a small longitudinal cross-section from $x$ to $x + \delta$, transverse fluid displacement occurs at a rate of approximately $\delta b(x) \dot{w}(x)$, as transverse fluid motion is generated only by the motion of the OCC and $\delta b(x)$ is the approximate area of the OCC in this range. 

Longitudinally, fluid enters the region at rate $S(x) \dot{u}(x)$ and exits at rate $S(x+\delta) \dot{u}(x+\delta)$. Exiting fluid must be the sum of entering fluid and fluid displaced by transverse OCC motion:
$$S(x+\delta)\dot{u}(x+\delta) = S(x)\dot{u}(x) + \delta b(x)\dot{w}(x).$$
This can be manipulated into the form of a difference quotient
$$\frac{S(x+\delta)\dot{u}(x+\delta) - S(x)\dot{u}(x)}{\delta} = b(x)\dot{w}(x).$$
Letting $\delta \rightarrow 0$, the left-hand side is recognized as a derivative in $x$:
\begin{equation}
    \frac{\partial}{\partial x} [S \dot{u}] = b\dot{w}.
    \label{consmassHorn}
\end{equation}

This identity can be used to arrive at a differential equation in pressure. This begins with simplifying the Navier-Stokes equation (Eqn \ref{navier}) to 1-D,  multiplying it by the cross-sectional area and differentiating in $x$:
\begin{equation}
    \frac{\partial}{\partial x} \bigg[ S\frac{\partial P}{\partial x}\bigg] + \rho \frac{\partial}{\partial x} \bigg[S \frac{\partial \dot{u}}{\partial t}\bigg] =0,
\end{equation}
where I have used the fact that $\dot{u} = \partial \phi/\partial x$.

Replacing the time derivative by product with $j\omega$ and applying Eqn \ref{consmassHorn} gives
\begin{equation}
    \frac{\partial}{\partial x} \bigg[S\frac{\partial P}{\partial x}\bigg] + j\omega\rho b \dot{w} =0.
\end{equation}
To write this entirely in terms of transmembrane pressure $p$, I can use $p=2P$ and the 1-D model's point-impedance boundary condition $p = -Z_{OC} \dot{w}$. Doing so and dividing by $S$ gives the Webster horn equation for the 1-D model:

\begin{align}
        \frac{1}{S}\frac{\partial}{\partial x} \bigg[S\frac{\partial p}{\partial x}\bigg] + k^2 p = 0, \label{webster1D}\\
        k^2(x) = \frac{-2 j\omega \rho}{Z_{OC}(x) h(x)}.
        \label{webster1Dk}
\end{align}

This derivation can be readily modified to apply to the 2-D model as well -- one replaces the 1-D model's longitudinal velocity $\dot{u}$ and pressure $P$ with those of the 2-D model averaged across the transverse dimension, and replaces the 1-D model's transverse velocity $\dot{w}$ with that of the 2-D model at $z=0$. The derivation holds identically until the final step.  In the 2-D model, the boundary condition is that transverse OCC velocity is related to the pressure at $z=0$, not the average pressure. Writing the average pressure as $\bar{P}$ (or $\bar{p} = 2\bar{P}$), the \textit{pressure-focusing  factor} is defined as 
\begin{equation}
\alpha(x) = \frac{p(x,0)}{\bar{p}(x)},  
\end{equation}
the ratio between the pressure focused at the OCC and average transmembrane pressure in the cross-section. The 2-D model Webster horn equation is then

\begin{align}
        \frac{1}{S}\frac{\partial}{\partial x} \bigg[S\frac{\partial \bar{p}}{\partial x}\bigg] + k_{2D}^2 \bar{p} = 0, \label{webster2D}\\
        k_{2D}^2(x) = \frac{-2 j\omega \rho \alpha(x)}{Z_{OC}(x) h(x)} .
        \label{webster2Dk}
\end{align}

In the box model where $S$ is constant, Eqns \ref{webster1D} and \ref{webster2D} degenerate to wave equations with variable wavenumbers, and Eqns \ref{webster1Dk} and \ref{webster2Dk} are dispersion relations.
\vfill
\pagebreak
\clearpage
\section{WKB Solutions for The 1-D Model}
\label{sec:1d}

The Webster horn equation (Eqn \ref{webster1D}) was studied in the context of cochlear mechanics models as early as 1950 \cite{Peterson_Bogert_1950}. For general values of $k(x)$, the PDE may not have a simple closed-form analytic solution, but approximations such as constant $k$ and simple forms for $S$ can be used to yield explicit solutions \cite{Peterson_Bogert_1950}. In the box model where this simplifies to a wave/transmission line equation with varying wavenumber, an explicit solution is guaranteed but only in terms of {retarded Green's functions\cite{Poisson_2004,Allen_1977,Allen_Sondhi_1979}\footnote{{Retarded Green's functions are tools used in the study of general relativity that are often challenging to write in closed-form. These are not to be confused with standard Green's function methods used in cochlear mechanics research, which generally give implicit solutions.}} . This motivates the development of a closed-form, explicit approximate solution.}

The WKB approximate solution for the horn equation\cite{hornwkb, hornwkb2}\footnote{{Briefer derivations of the WKB approximate solution to the horn equation appear in literature\cite{hornwkb, hornwkb2}, but are generally less detailed than the presentation in this tutorial.}} can be considered by putting the equation into the standard form of a linear differential equation:
\begin{equation}
    \frac{\partial^2 p}{\partial x^2} + \frac{S'}{S} \frac{\partial p}{\partial x} + k^2 p = 0,
    \label{hornLinearDE}
\end{equation}
where $\cdot'$ denotes the spatial derivative. Comparing with Eqn \ref{WKBODE}, we have $\epsilon = 1$. 

Plugging in the WKB ansatz (Eqn \ref{WKB_withdelta}) with $\delta = \epsilon = 1$ gives
\begin{equation}
    \exp{\bigg(\sum_{m=0}^\infty C_m \bigg)}\;\bigg[ \sum_{m=0}^\infty C_m'' + \bigg(\sum_{m=0}^\infty C_m' \bigg)^2 + \frac{S'}{S} \sum_{m=0}^\infty C_m' + k^2\bigg] = 0.
\end{equation}
As the exponential term is never 0 and leads every term, I can divide through by it. I choose to keep only terms involving $C_0$ and $C_1$. By the asymptotic assumptions of the WKB approximation $C_1 \ll C_0$, and the terms should decrease as further derivatives are taken so that $C_1 ''$ and $(C_1')^2$ {are negligible in comparison to lower-order terms}. This gives
\begin{equation}
    C_0'' + (C_0')^2 + 2C_0'C_1' + \frac{S'}{S} (C_0' + C_1') = -k^2.
    \label{highest}
\end{equation}

{To arrive at approximate solutions, even further simplifying assumptions must be made. At first these simplifications may appear as ``hand waving" for the sake of mathematical convenience, but they will be kept track of and analyzed at the end of the present section.}

{The most stringent approximation made in these derivations is the \textit{strong area assumption}
\begin{equation}
    \bigg|\frac{S'}{S} C_0'\bigg|\ll |C_0'|^2,
    \label{S_assumption}
\end{equation}
an assumption about the physical parameters of the system that can be interrogated once a first approximation of $C_0$ has been found. This is trivially true in a box model no matter the value of $C_0$, as in this case $S'=0$.}

{A consequence of this strong area assumption is the \textit{weak area assumption}, 
\begin{equation}
    \bigg|\frac{S'}{S} C_1'\bigg| \ll \bigg|\frac{S'}{S} C_0'\bigg|\ll |C_0'|^2,
    \label{S_assumption_1}
\end{equation}
justified by the asymptotic assumptions on the WKB series. Application of this weak area assumption to Eqn \ref{highest} gives the first-order WKB ODE:}
\begin{equation}
    C_0'' + (C_0')^2 + 2C_0'C_1' + \frac{S'}{S} C_0' = -k^2.
    \label{WKB1D_ODE1}
\end{equation}

{To obtain the zeroth-order WKB ODE, I perform a further reduction based in the same asymptotic decay assumptions. First, I make the \textit{second derivative assumption}
\begin{equation}
    |C_0''| \ll |C_0'|^2,
    \label{secderassumption}
\end{equation}
so that I am justified in ignoring the first summand of Eqn \ref{WKB1D_ODE1}. Ignoring also the remaining first order term of $2C_0'C_1'$ and applying the strong area assumption}, Eqn \ref{WKB1D_ODE1} reduces to 
\begin{equation}
    (C_0')^2 = -k^2.
    \label{WKB1D_ODE0}
\end{equation}
With Eqns \ref{WKB1D_ODE0} and \ref{WKB1D_ODE1}, zeroth- and first-order WKB approximations for the 1-D model can be found.

\subsection{The Zeroth-Order Solution}

The ODE in Eqn \ref{WKB1D_ODE0} is quickly solved by taking the square root of both sides and integrating:
\begin{equation}
    C_0 = \pm j \int_0^x k(\xi)\;d\xi.
    \label{WKB1d_C0}
\end{equation}
Plugging into the WKB ansatz, the zeroth order solution is
\begin{equation}
    p_0(x) = Ae^{-j\int_0^x k(\xi)\;d\xi} + Be^{j\int_0^x k(\xi)\;d\xi},\;\;\; A,B\in\mathbb{C}.
\end{equation}
This is recognized as a superposition of two traveling waves, with the first summand traveling towards the apex and the second summand traveling towards the base. As stated in Sec \ref{sec:conditions}, the basal-traveling waves due to reflection at the helicotrema are modeled as being negligible, so $B=0$.

Application of the boundary condition at the oval window (Eqn \ref{owBV}) with $B=0$ gives $A=P_{OW}$. The zeroth order solution is then 

\begin{equation}
    p_0(x) = P_{OW}e^{-j\int_0^x k(\xi)\;d\xi}.
    \label{WKB1Dpressure0}
\end{equation}

\subsection{The First-Order Solution}

Plugging in the value for $C_0$ found in Eqn \ref{WKB1d_C0} to the first-order WKB approximate ODE of \ref{WKB1D_ODE1}, I have
\begin{equation*}
    \pm j k' - k^2 \pm 2 k C_1' \pm \frac{S'}{S} jk  = -k^2.      
\end{equation*}
Solving for $C_1'$ gives
\begin{equation}
    C_1' = -\frac{k'}{2k} - \frac{S'}{2S},      
\end{equation}
and integrating both sides,
\begin{equation}
    C_1 = -\frac{1}{2}\log{(Sk)}.       
\end{equation}

Plugging the found values of $C_0$ and $C_1$ into the WKB ansatz gives the first-order WKB approximate solution,
\begin{equation}
    p_1(x) = \frac{A}{\sqrt{S(x)k(x)}}e^{-j\int_0^x k(\xi)\;d\xi} + \frac{B}{\sqrt{S(x)k(x)}}e^{j\int_0^x k(\xi)\;d\xi},\;\;\; A,B\in \mathbb{C}. \label{forback}
\end{equation}
Once again, $B=0$ by the assumption that basal-traveling waves are negligible. $A$ is found by applying the boundary condition at the oval window, and writing $S(0) = S_0, k(0) = k_0$:
\begin{equation}
    p_1(x) = P_{OW}\sqrt{\frac{S_0 k_0}{S(x)k(x)}}e^{-j\int_0^x k(\xi)\;d\xi}.
    \label{WKB1Dpressure1}
\end{equation}
In the box model, as $S(0) = S(x)$ for all $x$, the ratio inside the square root simplifies to $k_0/k(x)$. 

Eqns \ref{WKB1Dpressure0} and \ref{WKB1Dpressure1} are explicit formulae for pressure in terms of model parameters -- geometry and wavenumber, found through density, frequency and impedance (Eqn \ref{webster1Dk}). Having solved for pressure, velocity of the OCC in the 1-D model can be determined simply by dividing by the negative of the impedance.

\subsection{The Slowly-Varying Parameter Approximation}

The term \textit{WKB assumption} is often used to refer to the assumption that the wavenumber varies slowly in space relative to its own magnitude \cite{Zweig_compromise,Zweig_1991}. However, the derivation above for first- and second-order WKB approximations never explicitly made this assumption. Where is the relationship between these two ideas?

For the WKB method to be valid, the terms $C_n$ in the series must decrease monotonically. In particular $|C_1|\ll |C_0|$. Using the box model case for the sake of simplicity (no dependence on $S$), consider this relationship with the values of $C_0$ and $C_1$ from the 1-D model derived above, 
\begin{equation}
    \bigg|-\frac{1}{2}\ln{k}\bigg| \ll \bigg|j\int_0^x k(\xi)\;d\xi\bigg|.
\end{equation}

The left-hand expression can also be written as an integral from $0$ to $x$, and pulling out the constant-modulus factors gives
\begin{equation}
    \frac{1}{2}\bigg|\int_0^x \frac{k'(\xi)}{k(\xi)}\;d\xi\bigg| \ll \bigg|\int_0^x k(\xi)\;d\xi\bigg|.
\end{equation}

This relationship is satisfied if $k$ satisfies 
\begin{equation}
    |k'| \ll |k|^2.
    \label{wkbassume}
\end{equation}
That is, the assumption of slow-varying $k$ implies that the WKB approximation is reasonable. {This can also be recognized as the second derivative assumption (Eqn \ref{secderassumption}) made above in the derivation of the zeroth order WKB ODE.}.

{With the dependence of $k$ on the scala height and the impedance at the OCC (Eqn \ref{webster1Dk}), its rate of change is related to those of all of the model parameters. Thus, the} WKB assumption can be written as

\textit{\textbf{WKB Assumption:} The parameters of the model vary slowly relative to their magnitudes.}

Where these conditions are not met, the WKB approximation breaks down. It is important to keep this assumption in mind when observing modeled responses.

Consideration of asymptotic behavior of the cochlea's traveling wave in light of Eqn \ref{wkbassume} is instructive. At positions far basal to the best place, the response is said to be in the \textit{long-wave} region. Here, the wavenumber varies slowly in space, so the left-hand term in Eqn \ref{wkbassume} is very small.

At positions near the best place, the wavelength becomes smaller (wavenumber becomes larger) and changes more rapidly in space. This is known as the \textit{short-wave} region. Here, the right-hand term in Eqn \ref{wkbassume} is very large. In balance, this assumption may be satisfied across a large portion of the frequency range.

On the other hand, reasonable smooth values for impedance will lead to quickly varying $k$ around the region where stiffness and mass terms of the impedance cancel (see the dispersion relation of Eqn \ref{webster1Dk}). In lossless cases, this leads to an infinite admittance, and with small resistance still yields rapidly varying $k$.
\subsection{{The Strong Area Assumption and the Cochlear Catastrophe}}
{Further interrogation of the strong area assumption of Eqn \ref{S_assumption} is in order, as it also regards the spatial variation of model parameters. Under the first approximation for $C_0$ (Eqn \ref{WKB1d_C0}), the assumption becomes
\begin{equation}
    \bigg|\frac{S'}{S}\bigg| \ll |k|.
\end{equation}
This assumption appears to be challenged at the base, where $|S'/S|$ may be large as} scala area varies approximately exponentially \cite{plassmann}. 

{The presence of the wavenumber in this rewritten strong area assumption implies that there must be some balance between $k$ and $S$ to maintain the validity of the WKB approximation across the length of the cochlea. Reciprocally, appearance of the flattened scala height $h$ in the formula for the wavenumber (Eqn \ref{webster1Dk}) implies that the derivative of scala area also impacts the WKB assumption for wavenumber variation in Eqn \ref{wkbassume}.}  Zweig and Shera discuss the implications of this balancing act between geometry and OCC impedance in detail, and refer to the failure of models to account for this as the ``cochlear catastrophe" \cite{catastrophe}. It is worth noting, however, that this catastrophe is only significant in the base in response to very low-frequency stimuli, so box models without tapering reasonably satisfy the WKB approximation across most of space and frequency.

\pagebreak
\clearpage
\section{WKB Solutions for The 2-D Model}
\label{sec:2d}

There are a number of methods for arriving at WKB approximate solutions for the 2-D model. The first solution presented in this tutorial is chosen due to its emphasis of the relationship between the 1-D and 2-D models. Under the assumptions of the model (see Sec \ref{sec:3d}), transmembrane pressure satisfies the 2-D Laplace equation:
\begin{equation}
    \frac{\partial^2 p}{\partial x^2} + \frac{\partial^2 p}{\partial z^2} = 0.
\end{equation}
One classical method for solving the Laplace equation is separation of variables, where it is assumed that the transmembrane pressure can be written as a product of a function of only $x$ and a function of only $z$:
$$p(x,z) = \mathcal{X}(x)\mathcal{Z}(z).$$

If separation of variables were satisfied, the solution would be a linear combination of eigenfunctions with eigenvalue $k$. These are of the forms
\begin{align}
    p_{k} &= (A\cosh{k x} + B\cosh{k x})(Ce^{jk z} + De^{-jk z}), \label{coshxexpz}\\
    p_{k} &= (Ae^{jk x} + Be^{-jk x})(C\cosh{k z} + D\sinh{k z}).
    \label{coshzexpx}
\end{align}
We know that the $x$-dependence of the solution should have the form of a wave, so $\mathcal{Z}$ should have the hyperbolic trigonometric form seen in Eqn \ref{coshzexpx}:
$$\mathcal{Z}(z) = C\cosh{kz} + D\sinh{kz}.$$

Plugging in the boundary condition at the outer wall (Eqn \ref{wallBV}) gives
$$\mathcal{Z}'(h) = k[C\sinh{kh} + D\cosh{kh}] = 0,$$
yielding the relationship $C = -D/\tanh{kh}$. A hyperbolic trigonometric identity gives
\begin{equation}
    \mathcal{Z}(z) = \frac{D}{\sinh{kh}}(\sinh{kh}\sinh{kx} - \cosh{kh}\cosh{kx}) = \frac{-D}{\sinh{kh}}\cosh[k(z-h)]. 
\end{equation}
Separation of variables has already been broken, as $h$ depends on $x$, but this nonetheless gives motivation for writing the form of the solution as 
$$p(x,z) = \cosh{[k(z-h)]}\mathcal{X}(x),$$
where $k$ may also vary in $x$. In this form, the solution satisfies the boundary condition at the outer wall. 

Here I will make the first of two WKB approximations by assuming 1) the form of $\mathcal{X}$ is that of Eqn \ref{WKB_withdelta} with $\delta = 1$, and 2) {the WKB assumption (parameters vary slowly in $x$ relative to their own magnitudes) holds} so that $x$-derivatives of $\cosh[k(z-h)]$ are small. The Laplace equation to zeroth order becomes:
$$C_0^2\cosh[k(z-h)]\mathcal{X}(x) + k^2\cosh[k(z-h)]\mathcal{X}(x) = 0,$$
where the second term is the second $z$-derivative of $p$. Just as in the 1-D case, this gives
$$C_0 = \pm j\int_0^x k(\xi)\;d\xi.$$
Assuming that the basal-traveling wave is negligible, the pressure is 
\begin{equation}
    p(x,z) = A(x)\cosh{[k(z-h)]}e^{-j\int_0^x k(\xi)\;d\xi},
    \label{pwkb2d1}
\end{equation}
where $A(x)$ is some unknown function.

A dispersion relation is still needed so that $k$ can be solved for. To do so, recall the boundary condition at $z=0$ (Eqn \ref{bvImpedance}). Pressure and velocity potential are related by $p=-2\rho j\omega \phi$ (Eqn \ref{ptophi}), so their first derivatives in $z$ give
$$ \dot{w} = \frac{- 1}{2\rho j \omega} \frac{\partial p}{\partial z},$$
or plugging in Eqn \ref{pwkb2d1},
\begin{equation}
    \dot{w} = \frac{- 1}{2\rho j \omega} A(x) k \sinh{[k(z-h)]}e^{-j\int_0^x k(\xi)\;d\xi}.
\end{equation}

At $z=0$, the ratio of velocity and pressure is the negative of the OCC admittance, $-Y_{OC}$.  That is, 
\begin{align*}
    -Y_{OC} &= \frac{-1}{2j \omega \rho}\frac{ -A(x) k \sinh{[kh]}e^{-j\int_0^x k(\xi)\;d\xi}}{A(x) \cosh{[kh]}e^{-j\int_0^x k(\xi)\;d\xi} } \\
    &= \frac{k \tanh{[kh]}}{2j\omega \rho},
\end{align*}
giving the dispersion relation
\begin{equation}
    k\tanh{kh} = -2j\omega\rho Y_{OC}.
    \label{dispersion}
\end{equation}
This dispersion relation\footnote{\textbf{A note on terminology: }This equation is often called the \textit{eikonal equation} in literature due to an analogy to geometric optics, but this language is imprecise. Eikonal equations are a class of differential equations that appear elsewhere in the above derivations, and this algebraic expression is not such an equation. Instead, I will simply use the term \textit{dispersion relation}, as it is more descriptive and precise.} is transcendental and does not possess a unique solution for $k$ (the implications of this will be covered in detail in Sec \ref{sec:implement}). Eqn \ref{dispersion} is also independent of $A$, meaning it will be valid for any approximation of $p$ in the form of Eqn \ref{pwkb2d1}.

Solving Eqn \ref{dispersion} for impedance gives
$$Z_{OC} = j\omega \frac{-2 \rho}{k\tanh{kh}}.$$
This allows for an attractive interpretation of the impact of the impedance on the traveling wave. Due to the leading $j \omega$, this appears similar to a mass. In particular, defining the \textit{effective height} of the fluid as
\begin{equation}
    h_{e}(k) = \frac{1}{k\tanh{kh}},\;\;\;Z_{OC} = -2j\omega \rho h_{e},
    \label{eqn::heff}
\end{equation}
the impedance at the OCC is that of a column of fluid with this effective height.  {It should be noted that while this analogy is useful, both this ``mass" and this ``height" are generally complex-valued, and only approximately real at lower frequencies (see discussion of the long-wave solution in Sec \ref{sec::shortlong}).}

\subsection{Pressure Focusing}

To determine $A(x)$ in Eqn \ref{pwkb2d1}, recall that the \textit{average} pressure in the 2-D model must satisfy the Webster horn equation (Eqn \ref{webster2D}). The average pressure is 
\begin{align}
    \begin{split}
    \bar{p}(x) &= \frac{1}{h(x)}\int_0^{h(x)} A(x)\cosh[k(z-h)]e^{-j\int_0^x k(\xi)\;d\xi}\;dz\\
    &= \frac{1}{k(x) h(x)}A(x)\sinh[k(x)h(x)]e^{-j\int_0^x k(\xi)\;d\xi}.
    \end{split}
    \label{barp}
\end{align}

The pressure focusing factor  $\alpha = p(x,0)/\bar{p}(x)$ is required to find $k_{2D}$ in Eqn \ref{webster2Dk}, and can now be found using Eqn \ref{pwkb2d1}:
\begin{equation}
    \alpha(x) = \frac{k(x)h(x)}{\tanh{[k(x)h(x)]}}.
    \label{focus}
\end{equation}
This is independent of $A$, meaning it will be valid for any $p$ in the form of Eqn \ref{pwkb2d1}.

Plugging this into Eqn \ref{webster2Dk}, $k_{2D}^2$ can be found to be
$$k_{2D}^2 = \frac{-2j \omega Y_{OC} \rho k h}{h\tanh{[kh]}},$$
but by Eqn \ref{dispersion}, this simplifies directly to 
$$k_{2D}^2 = k^2.$$
In Sec \ref{sec:1d}, zeroth- and first-order WKB approximations for solutions to the Webster horn equation were derived. This gives an approximate formula for average pressure by directly copying Eqn \ref{WKB1Dpressure1}:
\begin{equation}
    \bar{p}(x) = P_{OW} \sqrt{\frac{S_0 k_0}{S(x) k(x)}} e^{-j\int_0^x k(\xi)\;d\xi}.
\end{equation}
Equating this with the earlier expression for $\bar{p}$ in Eqn \ref{barp}, it is possible to solve for $A(x)$:
\begin{equation}
    A(x) = P_{OW} \frac{k(x) h(x)}{\sinh[k(x)h(x)]}\sqrt{\frac{S_0 k_0}{S(x) k(x)}}.
\end{equation}

Finally, after this second application of a WKB approximation, a 2-D equation for pressure has been derived:
\begin{equation}
    p(x,z) = P_{OW} \frac{k(x) h(x)}{\sinh[k(x)h(x)]}\sqrt{\frac{S_0 k_0}{S(x) k(x)}} \cosh{[k(x)(z-h(x))]}e^{-j\int_0^x k(\xi)\;d\xi}.
    \label{WKB2Dpressure}
\end{equation}
In conjunction with the dispersion relation of Eqn \ref{dispersion}, this allows solution for pressure or velocity throughout the scala. 

\subsection{A Higher-Order 2-D Approximation}
\label{sec::higherorder}
The above derivation arrives at Eqn \ref{WKB2Dpressure} through two consecutive applications of the WKB approximation, and neatly piggy-backs off of 1-D results for average pressure. However, this formula is not the only solution referred to in literature as ``the WKB solution" for a 2-D model.

Various alternate approximation methods arrive at the following equation for pressure in a box model:
\begin{align}
    \begin{split}
    p(x,z) = P_{OW}\frac{k_0 h}{\cosh{[k(x)h]}\tanh{[k_0 h]}} \sqrt{
    \frac{\tanh{[k_0h]} + k_0h\,\text{sech}^2{[k_0h]}}{\tanh{[k(x)h]} + k(x)h\,\text{sech}^2{[k(x)h]}}} \;\;\times\\ \times\;\;\cosh{[k(x)(z-h)]} e^{-j\int_0^x k(\xi)\;d\xi }
    \label{duifhuisKING}
    \end{split}
\end{align}
(e.g. \cite{steele_lagrange,viergever_Book,duifhuis_moh}). It is clear that this has the form of Eqn \ref{pwkb2d1}, which means that this solution shares the same effective height, dispersion relation and pressure focusing factor as the solution derived above (Eqns \ref{eqn::heff}, \ref{dispersion}, \ref{focus}). 

One derivation of this formula involves the solution for $p$ as a formal power series approximation \cite{viergever_Book}, inspired by the physics of surface waves \cite{keller}. It is informative but lengthy, and an outline can be found in App \ref{app:series}. A second derivation of this formula follows from considering the Euler-Lagrange equations in a lossless box model (i.e. $Z_{OC}$ purely imaginary) \cite{steele_lagrange}. Neither derivation explicitly relies on the WKB \textit{approximation}, although they do rely on the WKB \textit{assumption} and feature the characteristic WKB phase term (the integral of the wavenumber). Intricate treatments of both derivations can be found at https://github.com/brian-lance/wkb-derivations. 

While Eqn \ref{duifhuisKING} behaves similarly to Eqn \ref{WKB2Dpressure}, its responses match numerical solutions better in the peak region. On the other hand, Eqn \ref{duifhuisKING} only holds for box models where $h$ is constant, not allowing for the modeling of cochlear tapering. Contemporary work is largely partial to the lower-order approximation of Eqn \ref{WKB2Dpressure} \cite{sisto_2021,sisto_2023,earhorn,altoe_2022,Shera_Altoe_2023}. Differences in the behavior between these solutions will be discussed in Sec \ref{sec:results}. 

\subsection{Long- and Short-Wave Solutions}
\label{sec::shortlong}
It is also instructive to consider the behavior of the solution in the long-wave (small $k$, basal to best place/lower frequency than best frequency) and short-wave (large $k$, near best place/best frequency) limits\cite{Peterson_Bogert_1950,Ranke_1950,Siebert_1973}\footnote{Notably, long- and short-wave solutions are not particular to the WKB approximation -- for historical long-wave solutions see the early work of Peterson and Bogert \cite{Peterson_Bogert_1950}, and for historical short-wave solutions see Ranke and Siebert.}. These approximations lie in the limiting behavior of the hyperbolic tangent for real argument $a\in\mathbb{R}$: 
$\tanh{a}\approx a$ if $a$ is small and $\tanh{a} \approx 1$ if $a$ is large.

The dispersion relations (Eqn \ref{dispersion}) in the long-wave and short-wave limits are
\begin{align}
    k_{lw}^2 = \frac{-2 j\omega \rho}{Z_{OC} h},\\
    k_{sw} = -2j \omega \rho Y_{OC}.
\end{align}
These are explicit solutions for the wavenumber in these regions, simplifying computation. Notably, $k_{lw}$ is precisely the wavenumber from the 1-D Webster horn equation (Eqn \ref{webster1Dk}). 

Considering the same limiting behavior for the pressure focusing factor (Eqn \ref{focus}) gives
\begin{align}
    \alpha_{lw} = 1,\\
    \alpha_{sw} = kh.
\end{align}
This reinforces the realization that the long-wave approximation and the 1-D solution are equivalent at $z=0$. The effective height from Eqn \ref{eqn::heff} also has corresponding long- and short-wave approximations:
\begin{align}
    h_{e,lw} = \frac{1}{hk^2},\\
    h_{e,sw} = \frac{1}{k}.
\end{align}

\begin{figure}
    \centering
    \includegraphics[width = .4\textwidth]{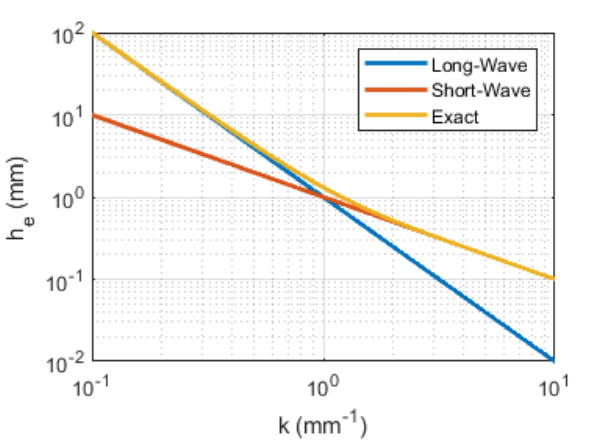}
    \caption{Effective height as a function of real $k$ for the long-wave and short-wave approximations, alongside the 2-D WKB solution. The scala height $h$ is set to 1 mm.}
    \label{fig::he}
\end{figure}

To visualize the difference between the long-wave, short-wave and WKB approximations, one can observe the effective height as $k$ varies. Fig \ref{fig::he} shows $h_e$ for the three solutions across various values of positive real $k$ with $h=1$ mm. It can be seen that the WKB solution for $h_e$ exhibits a continuous switch-off between the long- and short-wave approximations near the point where these solutions intersect. Behaviors of long- and short-wave velocity responses are discussed in Sec \ref{sec:results}.

\pagebreak
\clearpage
\section{The WKB Traveling Wave Subspace}
\label{sec:proj}

Secs \ref{sec:1d} and \ref{sec:2d} described derivations of explicit equations for pressure in 1- or 2-D tapered box models via the WKB approximation. These formulae are valid where the model parameters do not change rapidly relative to their magnitudes, and where basal-traveling waves are negligible. However, WKB approximate solutions may not be easily derived for other cochlear models.

Numerical solutions, while more accurate to the dynamics, are generally challenging to interpret in comparison to WKB approximate solutions. This problem arises in, for example, the study of reflections in cochlear models -- with only a numerical solution, how does one separate components of the response that are due to incident waves from those due to reflected waves? This same breakdown may also be challenging in solving alternate cochlear models featuring, for example, nonlinearity. It is in this context that the theory of \textit{cochlear basis functions} was developed \cite{basis}.

Consider a solution to a 1-D or 2-D cochlear model, describing the pressure at the OCC in response to a single-frequency stimulus. This is an infinite-dimensional object, living in the Hilbert space $\mathcal{H}$ of smooth functions mapping from $I=[0,L]$ (the interval of $\mathbb{R}$ along which the OCC is modeled to span) to $\mathbb{C}$. 

While the set of exact solutions is a subset of this infinite-dimensional space, they will likely have \textit{qualitatively} similar traveling wave forms for various perturbations to parameters and boundary conditions. Thus, they are likely to be well-approximated as living in a lower-dimensional subspace containing functions that resemble cochlear traveling waves. This motivates the concept of a \textit{traveling wave subspace}.

\subsection{Theory of Basis Wave Projection}

In a pioneering work, Shera and Zweig introduce several sets of basis functions that generate a traveling wave subspace, including the WKB basis functions \cite{basis}. I will develop the 1-D box model WKB basis, but the method is just as well extended to other approximate solutions. This specification is both for the sake of simplicity and because 1-D box model WKB basis functions are the most commonly seen in literature \cite{basis,oae_moleti,oae_sisto,Talmadge_Tubis_Long_Piskorski_1998,Talmadge_Tubis_Long_Tong_2000,Shera_Tubis_Talmadge_2005,Sisto_Moleti_Shera_2007}.

By Eqn \ref{WKB1Dpressure1} with constant cross-sectional area, the apical-traveling wave is proportional to
\begin{equation}
    W_+ = \sqrt{\frac{1}{k}}e^{-j\int_0^x k(\xi)\;d\xi}.
\end{equation}
The basal-traveling wave has been ignored thus far in this tutorial. However, Eqn \ref{forback} implies that the form of the WKB approximate solution for the basal-traveling wave would differ from $W_+$ only by the sign in the exponential. I define
\begin{equation}
    W_- = \sqrt{\frac{1}{k}}e^{j\int_0^x k(\xi)\;d\xi},\;\;\;x\in I.
\end{equation}

The set $\beta = \{W_+,W_-\}\subset \mathcal{H}$ is linearly independent, and its span, $\mathcal{W} = \text{\textbf{span}}(\beta)$ is a two-dimensional subspace of $\mathcal{H}$ which I will refer to as the \textit{WKB wave-space}. Any function $p \in \mathcal{W}$ can be written {in terms of its apical-traveling ($p_+$) and basal-traveling ($p_-$) components} as
\begin{equation}
    p(x) = p_+(x) + p_-(x) = \psi_+ W_+(x) + \psi_- W_-(x),\;\;\; x\in I,
    \label{p_proj1}
\end{equation}
where the coefficients $\psi_\pm$ are complex-valued constants. 

One can form a system of two equations in two variables by differentiating either side:
\begin{equation}
    \frac{\partial p}{\partial x} = \psi_+ \frac{\partial W_+}{\partial x} + \psi_- \frac{\partial W_-}{\partial x}.
    \label{p_proj2}
\end{equation}
Solution for the coefficients is neatly written in terms of the Wronskian of $\beta$\cite{Talmadge_Tubis_Long_Tong_2000,basis}\footnote{ {It should be noted that this Wronskian (also called the Wronskian determinant in some works) will be distinct for different choices of scaling factors for the basis functions. The projections will be identical.}}, which is
\renewcommand*{\arraystretch}{.7}
\begin{equation}
\mathcal{D} = \textbf{det}\begin{pmatrix}
            W_+ & W_- \\ W'_+ & W'_-
        \end{pmatrix} = 2j.
        \label{wronskian}
\end{equation}
With the Wronskian, the projections onto each basis function can be written as
\begin{align}
    \begin{split}
        p_\pm = \mathcal{P}_\pm [p] &= \psi_\pm W_\pm \\
        &= \frac{\pm W_\pm}{\mathcal{D}} \bigg(\frac{\partial W_\mp}{\partial x} - W_\mp \frac{\partial }{\partial x}\bigg) p \\
        &= \frac{1}{2}\bigg(1 \pm \frac{jk'}{2k^2} \pm \frac{j}{k}\frac{\partial}{\partial x}\bigg) p,
    \end{split}
    \label{projection}
\end{align}
with $\mathcal{P}_\pm$ representing the operators projecting functions in $\mathcal{H}$ onto $W_\pm$ \footnote{{Throughout this article, functions are written in terms of wavenumber $k$. Wavelength is used in other works\cite{basis}, causing a difference in the formulae presented here resolved simply by the chain rule of differentiation and the relationship between wavenumber and wavelength.}}\footnote{A more intricate treatment of these derivations can be found at https://github.com/brian-lance/wkb-derivations.}.

Of course, any exact solution to the BVP will not live in $\mathcal{W}$, so the values for $\psi_\pm$ found through this formula will not be constant. Thus, the projections are merely approximations that are best if the derivatives of $\psi_\pm$ are sufficiently small\footnote{A metric has been developed by Mathews and Walker and repeated by Shera and Zweig that quantifies the error through the size of these derivatives \cite{mathews_wkb,basis}. This is done by representing the basis functions as exact solutions to a similar differential equation $W'' + k^2(x)(1+\epsilon)W = 0$, with smaller $\epsilon$ representing better approximations to the actual BVP.}. 

Having these projection operators, one can formulate a numerical method for determining the apical- and basal-traveling components of any pressure waveform by implementing derivatives as finite differences. The same process can be followed for other basis functions of approximate solutions, such as the short-wave solutions,  long-wave solutions, or the WKB solutions in a tapered box model.

\subsection{Applications to Intracochlear Reflections}

One natural application of the projection described above is the study of reflections in the cochlea. The basal ($+$) and apical ($-$) reflection coefficients can be defined as 
\begin{equation}
    R_\pm(x) = \frac{p_\mp(x)}{p_\pm(x)}.
    \label{refdef}
\end{equation}
In a model of the cochlea where fluid pressure is driven at the stapes, a ``perfectly efficient" cochlea would reflect no energy in the basal direction (this is assumed in the derivations of Sec \ref{sec:1d} and Sec \ref{sec:2d}) and $R_+$ would be 0. In a cochlear model that exhibits some inefficiency, this will be a spatially varying complex-valued quantity. Some reasonable sources of such reflections include roughness in the OCC impedance or nonlinearity.

Conversely, one can consider how basal-traveling waves reflect towards the apex via $R_-$. With a passive cochlea driven at the stapes, this represents ``reflections of reflections." However, it is interesting to consider models where the cochlea is driven from a point along the length of the OCC ($x\neq 0$) \cite{basis,Talmadge_Tubis_Long_Piskorski_1998,Viergever_1986}. This could correspond to mechanical energy sources along the length of the OCC, present in the electromotile outer hair cells, which are likely responsible for many forms of OAEs. 

Given that OAEs are measurable when the cochlea is driven at the stapes, there must be some significant portion of energy traveling towards the base without being entirely reflected. Some early modeling work on this topic predicted that the apical reflection coefficient is very large compared to the basal reflection coefficient ($R_-\gg R_+$), so that basal-traveling energy would be significantly reflected before arriving back at the stapes \cite{Viergever_1986,de_Boer_Kaernbach_Konig_Schillen_1986}. In this formulation, OAEs would have very low magnitudes. It was later argued by Shera and Zweig that the sizes of these quantities are highly dependent on the boundary conditions of the model \cite{basis} -- an important result to keep in mind for the modeling of OAEs.

\subsection{Nonhomogeneous Models and WKB Solutions as a Fundamental Set}

The WKB basis functions may also be used as an analytic tool in finding approximate solutions to related cochlear models. Once again, I will specify to 1-D box models with constant area. Starting with Eqn \ref{webster1D}, the dynamics are governed by a wave equation with spatially-varying wavenumber
\begin{equation}
    \frac{d^2 p}{dx^2} + k^2(x) p = 0,
    \label{complementary_eqn}
\end{equation}
where I have replaced the partial derivatives with ordinary derivatives as time dependence is implicit. This is a second order linear homogeneous ODE for which $W_{\pm}$ are approximate, linearly independent solutions. If they were truly solutions, $\beta = \{W_+,W_-\}$ would form a \textit{fundamental set} for this ODE, and its general solution would be given by Eqn \ref{p_proj1}.

A corresponding nonhomogeneous ODE to Eqn \ref{complementary_eqn} is given by
\begin{equation}
    \frac{d^2 p}{dx^2} + k^2(x) p = g(x),
    \label{nonhomog}
\end{equation}
where $g$ is some non-zero function defined on $I$ known as the \textit{forcing function}. Eqn \ref{complementary_eqn} is the \textit{complementary equation} (or associated homogeneous equation) for this nonhomogeneous ODE. The theory of linear ODEs tells us that the general solution to Eqn \ref{nonhomog}, $p_{gen}$, can be written as the sum of the general solution to the complementary equation, $p_c$, and any particular solution to Eqn \ref{nonhomog}, $p_p$ \cite{Zill_Wright_Cullen_2013}. That is,
$$p_{gen} = p_c + p_p.$$

The complementary solution $p_c$ is a linear combination of the functions in the complementary equation's fundamental set, which can be approximated by the set of WKB solutions $\beta$. That is,
\begin{equation}
    p_c \approx a_+ W_+ + a_- W_-,\;\;\; a_\pm\in\mathbb{C}.
    \label{compl}
\end{equation}
The theory of variation of parameters \cite{Zill_Wright_Cullen_2013} then gives a particular solution in terms of the functions in the fundamental set and the forcing function:
$$p_p = \frac{W_-}{\mathcal{D}}\int_0^x W_+(\xi) g(\xi)\;d\xi - \frac{W_+}{\mathcal{D}}\int_0^x W_-(\xi) g(\xi)\;d\xi,$$
where $\mathcal{D}$ is once again the Wronskian of the WKB functions already found to be $2j$ in Eqn \ref{wronskian}.

This gives an approximate closed-form general solution for Eqn \ref{nonhomog}:
\begin{align}
    \begin{split}
        p_{gen} &= \bigg[a_+ -\frac{1}{2j}\int_0^x W_-(\xi) g(\xi)\;d\xi \bigg]W_+ + \bigg[ a_-  + \frac{1}{2j}\int_0^x W_+(\xi) g(\xi)\;d\xi\bigg] W_-\\
        &= p_+ W_+ + p_- W_-.
    \end{split}
    \label{gen}
\end{align}
The values of $a_\pm$ are found through the boundary conditions. In particular, if a known pressure is applied at the stapes ($x=0$) creating an initial apical-traveling wave, we would have $a_+= p_0$ (known constant) and $a_-$ = 0. This gives the parameter-free closed-form solution for the stapes-driven nonhomogeneous 1-D model:
\begin{equation}
    p = \bigg[p_0 -\frac{1}{2j}\int_0^x W_-(\xi) g(\xi)\;d\xi \bigg]W_+ + \bigg[\frac{1}{2j}\int_0^x W_+(\xi) g(\xi)\;d\xi\bigg] W_-.
    \label{stapesdrivensln}
\end{equation}

This formulation has broad applications in the modeling of intracochlear reflections and OAEs, where model equations can be manipulated into the form of Eqn \ref{nonhomog} \cite{Talmadge_Tubis_Long_Piskorski_1998,Talmadge_Tubis_Long_Tong_2000}. In these cases, the forcing function $g$ will generally represent sources of reflections such as random perturbations in impedance or nonlinearity. This interpretation is visible in Eqn \ref{stapesdrivensln}, where $g$ can be thought of as a kernel in the integral of the basis function traveling in the opposite direction of that for which it is a coefficient. That is, the size of the apical-traveling component is modulated by the basal-traveling wave weighted by $g$, and vice versa.

In the previous subsection, I discussed the application of projection onto WKB waves to approximating local reflection phenomena (Eqn \ref{refdef}). The application is natural in this analytic treatment as well, given the $p_\pm$ values in Eqn \ref{gen}.

\subsection{Example: Analytic Treatment of Roughness}

There are various applications of WKB basis functions to the study of cochlear phenomena (several described in \cite{Talmadge_Tubis_Long_Piskorski_1998,Talmadge_Tubis_Long_Tong_2000}). In particular, values of $g$ can be formulated to study sources of reflection, including nonlinear phenomena (e.g. distortion product otoacoustic emissions). Here, I will provide a representative and important example -- that of applying \textit{roughness} to the cochlea's parameters.

Much work has been done regarding the study of the impact of roughness on the impedance in generating intracochlear reflections \cite{Talmadge_Tubis_Long_Piskorski_1998,Talmadge_Tubis_Long_Tong_2000,Shera_Tubis_Talmadge_2005,oae_sisto,oae_moleti,Sisto_Moleti_Shera_2007}. That is, if the smooth impedance $Z_{s}$ were modified by a small longitudinally varying perturbation,
$$ Z(x) = Z_{s}(x) + \delta Z(x),$$
this would impact the wavenumber of the traveling waves in both directions (Eqns \ref{webster1Dk} in 1-D, \ref{dispersion} in 2-D). One could also model this as a roughening of the wavenumber, where the smooth wavenumber would be $k_s$ and the roughened (squared) wavenumber would be
$$k^2 (x)= k_{s}^2(x) + \delta k^2(x).$$
For example, $\delta k^2(x)$ may be modeled as samples from independent identically distributed normal distributions at each $x$. The roughness could also be designed to depend on stimulus frequency, but this dependence will be left implicit as it will not impact the derivations.

Rewriting the wave equation in terms of the roughened wavenumber, we have
$$\frac{d^2 p}{dx^2} + \big[ k_s^2(x) + \delta k^2(x)\big] p = 0,$$
which is in fact homogeneous and linear. However, $\delta k^2$ is not necessarily differentiable -- in fact, it ought not be as ``rough" implies non-smooth. This precludes use of the WKB approximation in its current form, as the WKB assumption (Eqn \ref{wkbassume}) is not well-posed.

Moving the $\delta k^2$ term to the opposite side gives
$$\frac{d^2 p_{r}}{dx^2} + k_s^2(x)  p = -\delta k^2 p,$$
which is still homogeneous as the right-hand side is proportional to the dependent variable $p$. The strategy is to approximate the right-hand side as a $p$-independent forcing function. If $\delta k^2$ is small, we can consider this right-hand term as a perturbation to the otherwise smooth, complementary response $p_c$ of Eqn \ref{compl}. In the case that an apical-traveling wave of magnitude $p_0$ is induced at the stapes, this gives the approximation 
$$-\delta k^2 p \approx -\delta k^2 p_0 W_+.$$
This is a known $p$-independent function, allowing the ODE to be interpreted as approximately nonhomogeneous. 

That is, it is in the form of Eqn \ref{nonhomog} with $g = -\delta k^2 p_0 W_+$. The roughened pressure solution can be given by substituting this forcing function for $g$ in Eqn \ref{stapesdrivensln}:
\begin{equation}
    p = p_0\bigg[1+ \frac{1}{2j}\int_0^x \delta k^2(
    \xi)W_+(\xi)W_-(\xi) \;d\xi \bigg]W_+ - p_0 \bigg[\frac{1}{2j}\int_0^x \delta k^2(\xi) W_+^2(\xi) \;d\xi\bigg] W_-.
    \label{roughresponse}
\end{equation}
This solution facilitates computation of the reflection coefficients through Eqn \ref{refdef} in terms of the roughness function\footnote{ {The formula is similar to Eqn 74 of \cite{Talmadge_Tubis_Long_Tong_2000}, save (1) a negative sign and (2) complementary integral bounds. These discrepancies cancel if the integral over the interval is 0.}}. 
\vfill
\pagebreak
\clearpage
\section{Implementation: Solving the Dispersion Relation}
\label{sec:implement}

In the previous sections I have described theoretical underpinnings for WKB solutions to 1-D and 2-D box and tapered box models. In this section, I discuss the challenges involved in implementation of the derived model equations in software. 

The 1-D model poses no such difficulty, as the WKB pressure equation and dispersion relation are explicit and in terms of elementary functions, but the dispersion relation of Eqn \ref{dispersion} presents a challenge in the 2-D case. This relation is transcendental, and generally has infinitely many solutions for $k$ in the complex plane. To standardize the language, the solution for $k$ is reframed as a root-finding problem for the function
\begin{equation}
    f(z) = z\tanh{z} - C,
    \label{rootfind}
\end{equation}
where
\begin{equation}
    z=kh,\;\;\;C=-2\rho h j\omega Y_{OC}.
    \label{Cdef}
\end{equation}

At each position and frequency, solutions will exist for multiple values for $k$, but we will select only the most significant of such modes\footnote{This \textit{one-mode assumption} is worthy of some scrutiny, and the works of Watts \cite{watts,Watts_2000} and Elliott \cite{Elliott_Ni_Mace_Lineton_2013} have covered multi-mode solutions and implications thereof.}. As the velocity is loosely of the form $e^{-jkx}$, the solution should possess a positive real part to correspond to {an apical-traveling wave. As for the imaginary part, this leads to either dampening or amplification of the solution in $x$. Exponential growth is not physical in a passive cochlea, meaning that the imaginary part must be negative and the solution for $k$ must lie in the $4^{\text{th}}$ quadrant of the complex plane.}

Moreover, of the solutions in this quadrant, the one with the smallest (in magnitude) imaginary part is desired. A more negative imaginary part would lead to more severe exponential damping, so the most significant solution has the least such damping.

In this section, I discuss the properties of the roots of $f$, and the challenges that come in solving for physically reasonable roots. I then describe in detail three algorithms for finding $k$. The performance of these algorithms is discussed in Sec \ref{sec:results}.

\subsection{Root Loci}

\begin{figure}[ht]
    \centering
    \includegraphics[width = .6\textwidth]{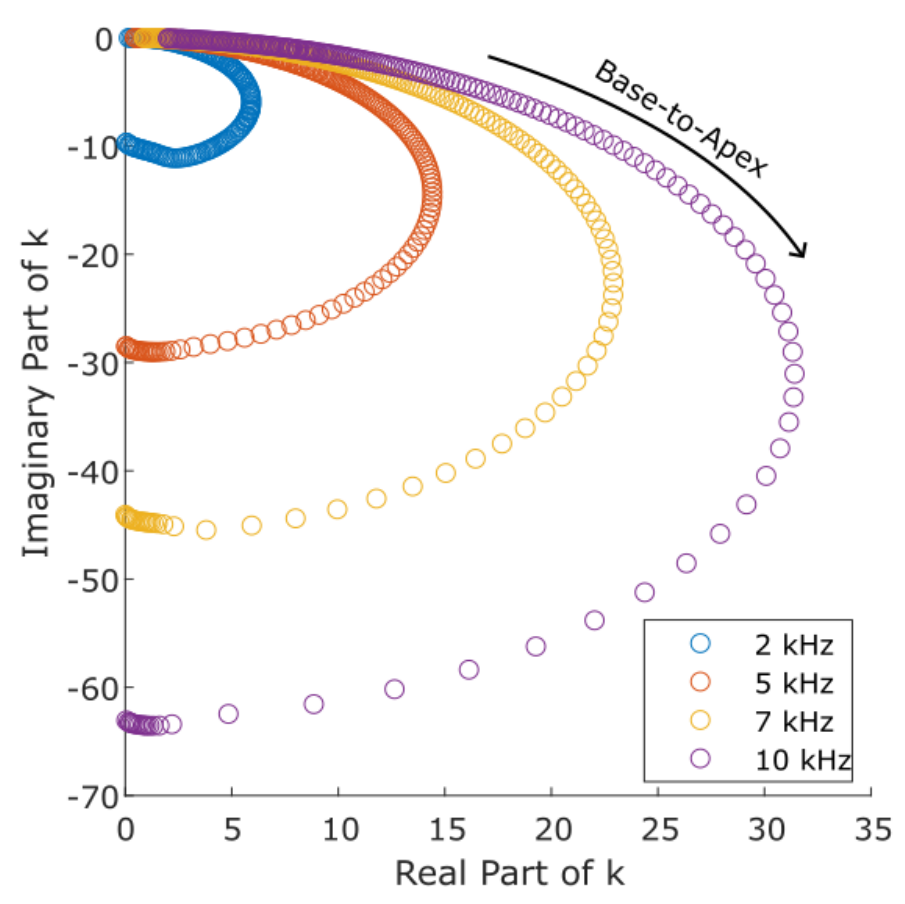}
    \caption{Root loci for $f(z)$ in response to four stimulus frequencies using parameters from Steele and Taber \cite{steele_lagrange} (see Table \ref{tab:params}). For a single color, the most basal position corresponds to the smallest real root. As $x$ increases in 7 $\mu$m increments, the root traverses a clockwise arc in the fourth quadrant, eventually arriving at the negative imaginary axis.}
    \label{fig:loci}
\end{figure}

Because the function $f$ is continuous, a small variation of $C$ should yield a small variation of the root position. Each continuous path traced by the roots with increasing $x$ is called a \textit{root locus}. With realistic impedance functions, the root loci form arcs in the fourth quadrant, traveling from the positive real axis to negative imaginary axis with increasing $x$ \cite{deBoer_Rootfinding}. Fig \ref{fig:loci} shows four such root loci under one set of parameters, where each color corresponds to a different stimulus frequency and each circle is a root at a different $x$ position ($x$-resolution is $7\mu m$). As $x$ increases, the locus diverges from the real line and traverses clockwise towards the negative imaginary axis. At higher frequencies, the arc is broader and arrives at a larger negative imaginary value.

The WKB assumption is that this variation of $k$ in $x$ is slow, so that tracing the continuous arc through the plane (possible with a fine enough resolution in $x$) would give the root of interest. However, with physically realistic parameters, one runs in to multiple issues just past the peak region. In particular, near the location where stiffness and mass of $Z_{OC}$ cancel (Eqn \ref{impedance}, Table \ref{tab:params}), the admittance factor of $C$ varies rapidly. Here, the WKB assumption breaks down, and a tracing of the root locus shows a rapid traversal of the arc near these positions. This can be seen in Fig \ref{fig:loci} where the roots appear sparse along the broad arc of the locus, indicating a much faster change in $k$ than at the denser regions near the real and imaginary axes. In this region, insufficient resolution in $x$ could not capture the continuous but rapid arc of the root locus, and may instead yield convergence to a root in a different locus. This issue can be resolved either by uniformly refining resolution, or refining resolution
close to the resonant position \cite{viergever_Book}. 

\subsection{Continuous Longitudinally-Stepping Algorithm}

The goal is to begin by tracing a single root locus for $f$ through the complex plane. Due to the number of possible roots at a given $x$, canonical root-finding methods can cause trouble. Such methods require an intelligently chosen starting point so as not to converge to the wrong root, or even a saddle point.

\par{In this section, I will describe a class of algorithms for root-finding that step across the longitudinal axis, at each point making an estimate for $k$ informed by the estimate from the previous step \cite{deBoer_Rootfinding,viergever_Book}. {Here, $x$ values are quantized into an $N$-length vector with resolution $\Delta x$}. I will write the estimate for $k$ at position $x_n$ as $\hat{k}_n$, $n=1,2,\ldots,N$. As the function $f$ is itself $x$-dependent, I will write $f(z;x_n)$ to refer to $f$ at each position.}

\par{Starting at the very base, we are likely to be in the long-wave region. This motivates the initial approximation of $\hat{k}_1 = k_{lw}(x_1)$. This can be used as the initial value in a standard root-finding algorithm such as Newton-Raphson or the Muller method, which are likely to converge to the correct root.}

\par{Stepping further along in $x$, the long-wave approximation becomes poor. This indicates that we ought not use this initial value forever. As in Fig \ref{fig:loci}, wavenumbers within a single locus at subsequent $x$ locations are likely close to one another -- that is, $k_n \approx k_{n+1}$. The intuitive estimate is to use the solution at $x_n$, $\hat{k}_n$, as the starting point for the root-finding method at $x_{n+1}$ to find $\hat{k}_{n+1}$.}

\par{Pseudocode for this algorithm using the Newton-Raphson method in $z$ to compute the wavenumber is presented in Alg \ref{alg::no_discon}. The Newton-Raphson method requires the derivative of $f$, given by
\begin{equation}
    f'(z) = \tanh{z}+z\text{sech}^2{z}.
\end{equation}
Recall that $z = kh$.}

\begin{algorithm}
\begin{algorithmic}
\State $\hat{k}_{ic} \gets k_{lw}(x_1)$ \Comment{Initialize using long-wave $k$}
\For{$n = 1 \rightarrow N$} \Comment{$N$ is the number of steps in $x$ space}
    \State{$z \gets h\hat{k}_{ic}$}
    \For{$m = 1 \rightarrow M$} \Comment {$M$ is the number of Newton-Raphson iterations}
        \State{$z \gets z - \frac{f(z;x_n)}{f'(z;x_n)}$} 
    \EndFor
    \State{$\hat{k}_{n} \gets z/h$}
    \State{$\hat{k}_{ic} \gets k_n$} \Comment{Initial value for next step is current guess for $k$}
\EndFor

\end{algorithmic}
\caption{Continuous longitudinally-stepping root-finding algorithm to determine the wave number at $N$ different $x$ positions, using the Newton-Raphson method.}
\label{alg::no_discon}
\end{algorithm}

This works so long as $k$ is slowly varying, which is precisely the WKB assumption. However, there is a significant problem that occurs near the resonant point where stiffness and mass cancel, creating a rapid change in $k$ (see the sparse regions of the arcs in Fig \ref{fig:loci})\footnote{\textbf{Note on terminology:} Some modelers have described this point as the \textit{critical layer}, owing its name to a more general theory of critical layer absorption \cite{lighthill_long,lighthill_short}.}. If we were to ignore this feature, we would simply follow the continuous root locus as in Fig \ref{fig:loci}, tending towards solutions for $k$ with large negative imaginary parts. This leads to falloff in the magnitude response that is far more rapid than what is seen in basilar membrane displacement data \cite{Robles_Ruggero_2001,steele_lagrange}.

{This rapid falloff past the peak is a result of a poor selection for $k$. In particular, }the roots along the continuous locus do not correspond to dominant modes once their imaginary parts become sufficiently negative. Methods have been developed to counteract this problem by considering a continuous switch-off between dominance of two modes \cite{watts,Watts_2000,Elliott_Ni_Mace_Lineton_2013}, or by discretely swapping the root locus being followed near the resonant position \cite{viergever_Book,steele_rootfinding}. An example of the latter type is discussed in the following subsection.

\subsection{Discontinuous Longitudinally-Stepping Algorithm}

To account for the {issues found in tracing the continuous root locus near the resonant position, one can introduce a discontinuity into the $x$-stepping algorithm. This is done by changing the initial value for the root-finding algorithm near the resonant position, where the WKB assumption breaks down. To do so, we seek a better guess for a root near this position.} 

To begin, the term $C$ in the root-finding problem (see Eqn \ref{Cdef}) is approximated near the resonant position. Where the stiffness and mass cancel, the admittance is $Y_{OC}\approx 1/R_d$, a real conductance (where the $d$ subscript denotes evaluation near the resonant position), i.e. $C$ is purely negative imaginary. I define a new term $\gamma$:
\begin{equation}
    \gamma = \frac{R_d}{2\rho h \omega}\in\mathbb{R}.
\end{equation}
The new problem to solve becomes
\begin{equation}
    z\tanh{z} = \frac{-j}{\gamma}.
    \label{gammaform}
\end{equation}

The solution to this transcendental equation can be approximated using the assumption that $z$ is small in magnitude and lives close to the imaginary axis. This ensures that the chosen value of $k$ will correspond to the dominant mode, falling off less rapidly than what would be found by tracing the continuous root locus. A derivation of this solution\footnote{An intricate treatment can be found at https://github.com/brian-lance/wkb-derivations}, relying on Taylor expansions, is given in brief by Viergever \cite{viergever_Book}. It yields
\begin{equation}
    k_d \approx \frac{\pi}{2h}\gamma - j \frac{\pi}{2h}(1-\gamma^2),\;\;\; \gamma = \frac{R_d}{2 \rho h \omega}.
    \label{kd}
\end{equation}

The discontinuous $x$-stepping algorithm traces the continuous root locus up to some $x_d$ at which it is determined that the WKB assumption is being violated. This can be determined before simulation \cite{viergever_Book} or on the fly by observing the rate of change of the wavenumber (in discrete space, the finite difference) at each step. When the WKB assumption holds, this rate should be small. Picking some threshold $T>0$, the $x$-stepping method is paused once {$|\hat{k}_n - \hat{k}_{n-1}|/\Delta x>T$}.

After this point, $k_d$ of Eqn \ref{kd} is used as an initial step in the root-finding algorithm. If still {$|\hat{k}_n - \hat{k}_{n-1}|/\Delta x>T$}, the WKB assumption is violated and the pressure at this position is set equal to its value at the last computed position ($p_n = p_{n-1}$). At each subsequent step, it is determined whether $\hat{k}$ satisfies this threshold -- it will eventually do so, at which point we continue the locus-tracing process along this second locus. Pseudocode for this algorithm is presented in Alg \ref{alg::with_discon}.

\begin{algorithm}
\begin{algorithmic}
\State{$\hat{k}_{ic} \gets k_{lw}(x_1)$} \Comment{$\triangleright$ Initialize using long-wave $k$}

\For{$n = 1 \rightarrow N$} \Comment{$\triangleright$ $N$ is the number of steps in $x$ space}
    \State{$z \gets h\hat{k}_{ic}$}
    \For{$m = 1 \rightarrow M$}   \Comment{$\triangleright$ $M$ is the number of Newton-Raphson iterations}
        \State{$z \gets z - \frac{f(z)}{f'(z)}$} 
    \EndFor
    \If{$\frac{|z/h - k_{ic}|}{\Delta x} < T$}   \Comment{$\triangleright$ Check for validity of WKB condition}
        \State{$\hat{k}_{n} \gets z/h$}
        \State{$\hat{k}_{ic} \gets k_n$}   \Comment{$\triangleright$ Initial value for next step is current guess for $k$}
    \Else{}
        \State{$\hat{k}_{n} \gets$ NaN}   \Comment{$\triangleright$ Pressure and velocity should not be computed here}
        \State{$\hat{k}_{ic} \gets k_d$}   \Comment{$\triangleright$ Guess for $k$ after the discontinuity}
    \EndIf
\EndFor

\end{algorithmic}
\caption{Longitudinally-stepping root-finding algorithm to determine the wave number at $N$ different $x$ positions, accounting for the discontinuity in the wavenumber. }
\label{alg::with_discon}
\end{algorithm}

\subsection{The Fixed Point Algorithm}

One alternative to the longitudinally-stepping class of algorithms is a fixed point algorithm, in which two distinct relationships between $k$ and $\alpha$ (the pressure-focusing factor) are used\cite{Shera_Tubis_Talmadge_2005,earhorn}. The first such relationship is Eqn \ref{focus}, which gives $\alpha$ in terms of $k$. The second, giving $k$ in terms of $\alpha$, is Eqn \ref{webster2Dk}. Any valid $k$ value must satisfy both equations.

\par{Fixed point methods are based on the Banach fixed point theorem \cite{Latif2014}, which states that repeated application of a contractive function $g$ will converge to a fixed point of said function, i.e. a point where $g(x) = x$. Mathematical details are omitted here for the sake of brevity.}

\par{The fixed point method for this problem works by starting with the long-wave approximation at every frequency-location pair, $\hat{k} = k_{lw}$. Then, $\hat{k}$ is plugged in to Eqn \ref{focus} to find an the approximate pressure focusing factor $\hat{\alpha}$, and then  $\hat{\alpha}$ is plugged in to Eqn \ref{webster2Dk} to find a new wavenumber approximation $\hat{k}$. This is repeated for some number of iterations. Pseudocode for this algorithm is shown in Alg \ref{alg::stable}. 
}

\begin{algorithm}
\begin{algorithmic}
\State{$\hat{k} \gets k_{lw}$}   \Comment{$\triangleright$ Here $\hat{k}$ is a vector with an index for each position}

\For{$m = 1 \rightarrow M$}    \Comment {$\triangleright$ $M$ is the number of fixed point iterations}
    \State{$\hat{\alpha} \gets \frac{h \hat{k}}{\tanh{kh}}$}  \Comment{$\triangleright$ Pressure focusing vector update, Eqn \ref{focus}}
    \State{$\hat{k} \gets \sqrt{\frac{-2j\omega \rho \hat{\alpha}}{h Z_{OC}}}$}   \Comment{$\triangleright$ Wavenumber update, Eqn \ref{webster2Dk}}
    \If{$\mathcal{R}[\hat{k}] < 0 $}
    \State{$\hat{k} \gets -\hat{k}$}  \Comment{$\triangleright$ Ensure the root is for an apical-traveling wave ($\mathcal{R}$ gives the real part)}
    \EndIf
\EndFor

\end{algorithmic}
\caption{Fixed point algorithm that updates a vector of $\hat{k}$ approximations by iteratively applying two relationships.}
\label{alg::stable}
\end{algorithm}

\par{This method works under the assumption that it converges to the correct value of $k$, which depends on the properties of the mappings between $\alpha$ and $k$. If the mappings are not (at least locally) contractive, then no convergence is guaranteed. On the other hand, if there are multiple fixed points, certain choices of initial conditions may lead to convergence to an undesired $k$. Performance of these three algorithms will be discussed in the following section.}

\vfill
\pagebreak
\clearpage
\section{Behavior of WKB Approximate Solutions}
\label{sec:results}

Having developed the WKB approximate solutions, as well as methods by which to find the wavenumber, it is now possible to observe the behavior of the modeled solutions. WKB solutions are compared to numerical results computed using the finite difference method of Neely \cite{neely}. Physical quantities used here are from the 2-D box model of Steele and Taber \cite{steele_lagrange}. These, along with parameters used in wavenumber-finding algorithms, are provided in Table \ref{tab:params}. Mass, resistance and stiffness terms contribute to the impedance according to 
\begin{equation}
Z_{OC}(x) = j\omega m + r + \frac{s(x)}{j\omega},
\label{impedance}
\end{equation}

where $x$ has units mm.

\renewcommand*{\arraystretch}{.5}
\begin{table}[h!]
    \centering
    \begin{tabular}{|c|c|c|}
        \hline Parameter & Symbol & Value \\
        \hline\hline Mass & $m$ & $1.5 \times 10^{-3}$ g/mm$^2$  \\
        \hline Resistance & $r$ & $2\times10^{-6}$ Ns/mm$^3$ \\
        \hline Stiffness & $s(x)$ & $10 e^{-0.2 x}$ N/mm$^3$ \\
        \hline Scala Height & $h$ & 1 mm \\
        \hline Cochlea Length & $L$ & 35 mm \\
        \hline Fluid Density & $\rho$ & $10^{-3}$ g/mm$^3$ \\ 
        \hline Threshold on $k$ finite difference & $T$ & 0.02 mm$^{-2}$\\
        \hline Iterations of Newton's Method (Algs \ref{alg::no_discon}, \ref{alg::with_discon}) & $M$ &  20 \\ 
        \hline Iterations of Contractive Mappings (Alg \ref{alg::stable}) & $M$ &  20 \\ 
        \hline Points in $x$-space & $N$ & 1024 \\
        \hline Points in $z$-space (finite difference method)& N/A & 16 \\ \hline
    \end{tabular}
    \caption{Parameters used in all simulations. Physical parameters are from Steele and Taber \cite{steele_lagrange}.}
    \label{tab:params}
\end{table}

\subsection{WKB Solutions for the 1-D Box Model}

\begin{figure}
    \centering
    \includegraphics[width=0.5\textwidth]{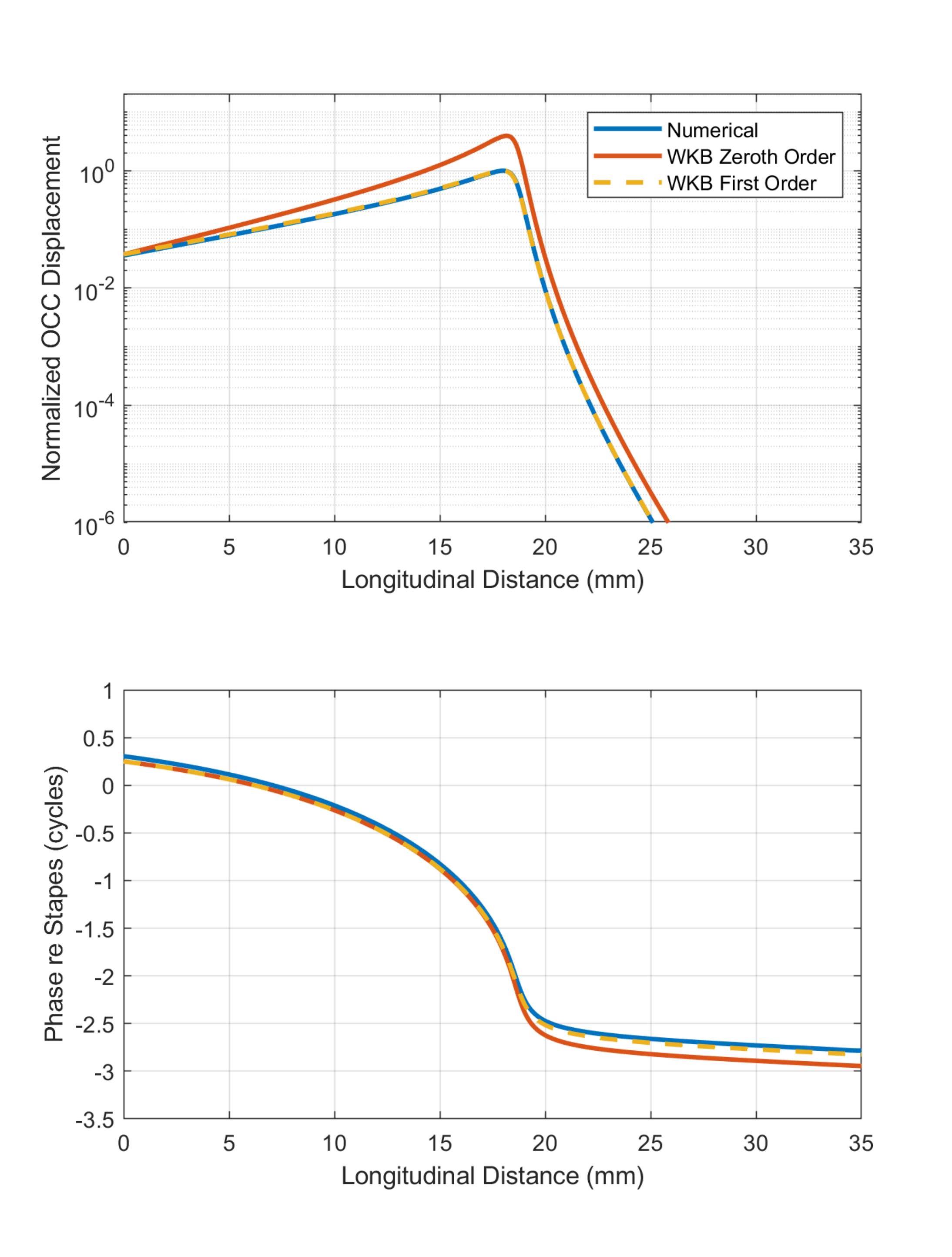}
    \caption{Comparison of numerical solutions to the 1-D box model with zeroth- and first-order WKB approximate solutions in response to a 2 kHz stimulus.}
    \label{fig:1d}
\end{figure}

In Sec \ref{sec:1d}, I derived WKB solutions to the 1-D BVP up to the zeroth (Eqn \ref{WKB1Dpressure0}) and first (Eqn \ref{WKB1Dpressure1}) orders. In Fig \ref{fig:1d}, these solutions are shown in response to a 2 kHz stimulus frequency, and compared to numerical results.

It can be seen that the zeroth-order approximation overestimates the magnitude of the response near the peak, and exhibits more phase accumulation than the numerical solution. On the other hand, the first-order WKB approximation matches the numerical solution remarkably well across space in both phase and magnitude. The two orders of solution differ only by a factor of $\sqrt{k}$, which is real-valued for small $x$ explaining the similarity in phase.

\subsection{Long-Wave and Short-Wave Solutions}

 To contextualize findings for the 2-D WKB solutions, it is useful to observe the performance of the long- and short-wave approximate solutions to the 2-D box model (see Sec \ref{sec::shortlong}, as well as  \cite{Peterson_Bogert_1950,Ranke_1950,Siebert_1973}). These solutions are valid for regions of small real $k$ and large real $k$ respectively, but as $k$ is complex-valued and varies non-monotonically across space/frequency (see Fig \ref{fig:loci}) it is not immediately clear in which regions these approximations will best match numerical solutions.

\begin{figure}
    \centering
    \includegraphics[width=0.5\textwidth]{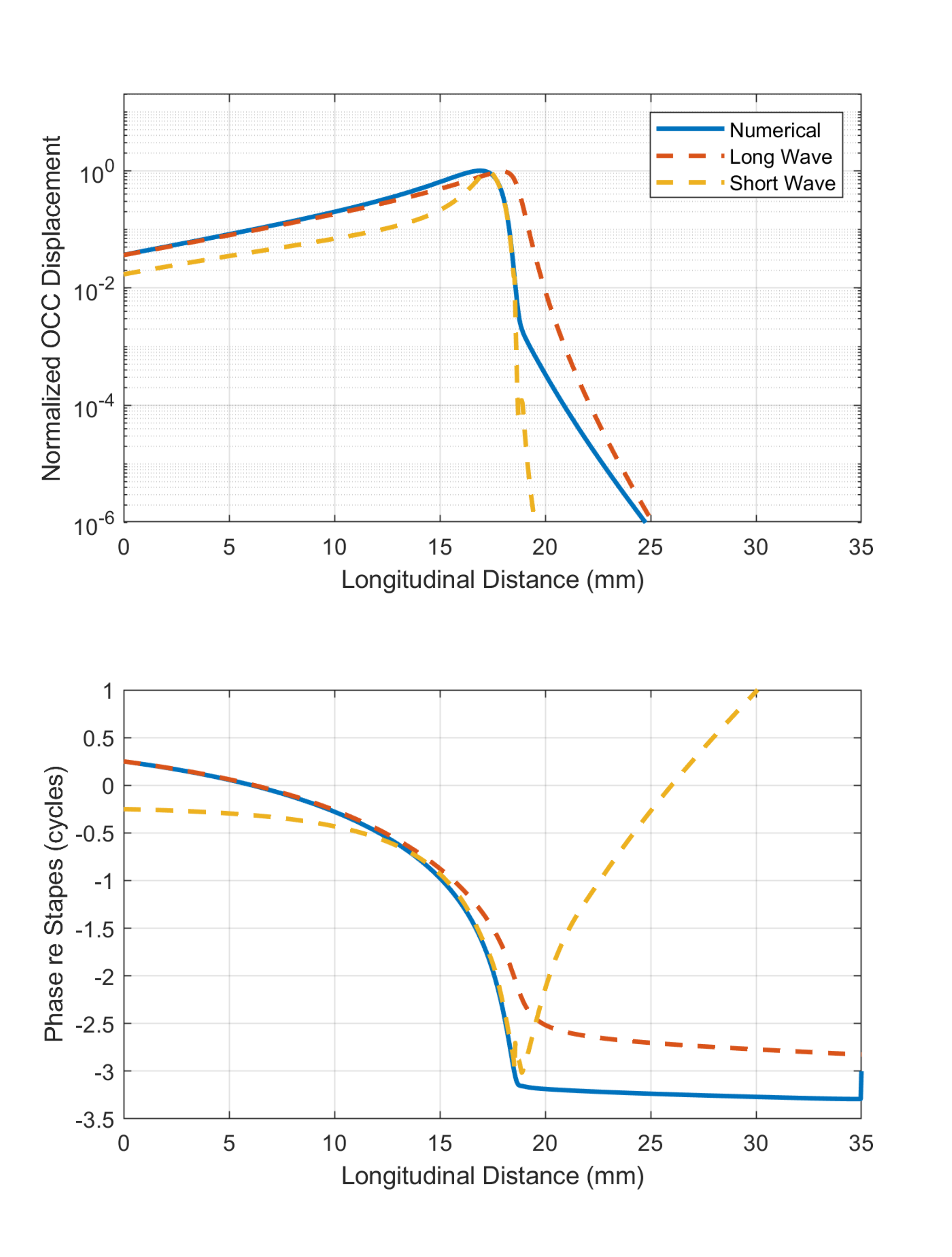}
    \caption{Comparison of numerical solutions to the 2-D box model with long- and short-wave approximate solutions in response to a 2 kHz stimulus.}
    \label{fig:lwsw}
\end{figure}

Fig \ref{fig:lwsw} shows the long-wave and short-wave solutions to the 2-D box model alongside a numerical solution. The long-wave response matches the numerical solution well at more basal positions, where the wavenumber is small and real (Fig \ref{fig:loci}), but poorly matches the numerical solutions near or above the peak.

The short-wave solution matches the numerical solution well only in a small spatial range near the peak. Neither approximation matches the numerical solution past the peak where the magnitude falloff in the numerical solution becomes slower. {This region is termed the \textit{cutoff region} \cite{watts}. In the context of the root loci (Fig \ref{fig:loci}), neither approximation should be expected to hold well in the cutoff region where the roots approach the negative imaginary axis, as the asymptotic forms of the hyperbolic tangent used in their derivations are only valid for real argument. The short-wave solution exhibiting a sign change in its group velocity just past the peak is one dramatic consequence of this breakdown.}

\subsection{Performance of Wavenumber-Finding Algorithms}

Before comparing 2-D WKB approximations to numerical results, it is first important to assess the methods for determining the wavenumber $k$ in the 2-D case. This is performed by observing velocity responses at the OCC derived from the 2-D WKB approximation of Eqn \ref{duifhuisKING}, using three methods for finding the wavenumber: 1) Alg \ref{alg::no_discon}, an $x$-stepping algorithm that does not account for the discontinuity, amounting to following a root locus as in Fig \ref{fig:loci}, 2) Alg \ref{alg::with_discon}, an $x$-stepping algorithm that does account for discontinuity, via thresholding the finite difference as described above, and 3) Alg \ref{alg::stable}, the fixed point method. 

%\par{In all cases the parameters of Steele and Taber are used \cite{steele_lagrange}, along with a 5.5 kHz stimulus. In the $x$-stepping methods, Newton-Raphson is applied for 20 iterations at each location, and resonance is accounted for in method (2) by using a threshold of $T=0.8$. In the stable point method, the iterative algorithm is applied 20 times.}

\begin{figure}[ht!]
    \centering
    \includegraphics[width = .7\textwidth]{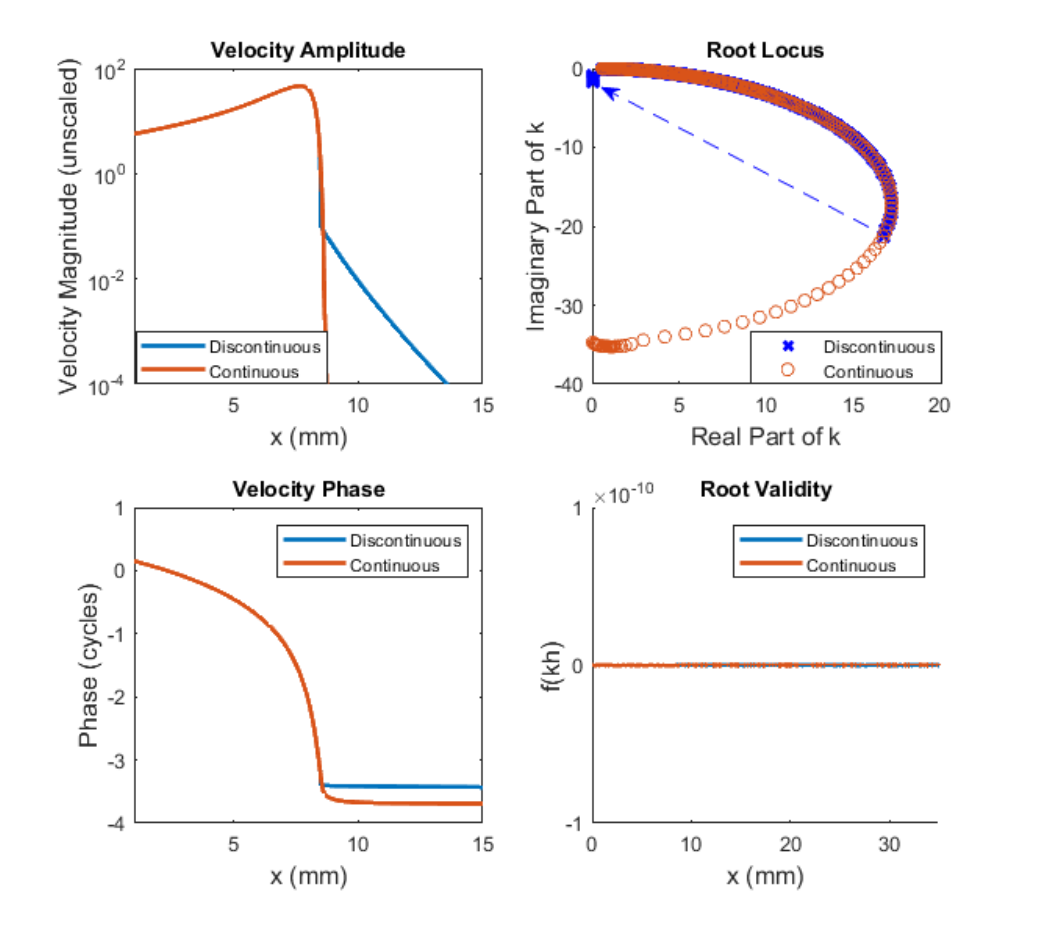}
    \caption{Simulated velocity of the OCC in response to a 5.5 kHz stimulus, using the parameters of Table \ref{tab:params}. Left-hand panels show the magnitude and phase responses for velocity computed using the 2-D WKB solutions, with the dispersion relation solved via $x$-stepping methods. The method labeled as ``Discontinuous" accounts for the resonance by stopping computations when the derivative exceeds a threshold (Alg \ref{alg::with_discon}), and the method labeled as ``Continuous" simply follows a continuous root locus (Alg \ref{alg::no_discon}). The followed loci are displayed in the top-right panel, with the dashed arrow indicating the jump in the discontinuous method when entering the cutoff region. The bottom-right panel shows $f(kh)$, which should be identically zero at roots. }
    \label{fig:disc}
\end{figure}

Fig \ref{fig:disc} contrasts the $x$-stepping methods depending on whether discontinuity is accounted for. The velocity responses show identical behavior up to a position slightly past the peak, where the finite difference in $k$ becomes sufficiently large so that a discontinuity is registered. After this point, the fall-off in velocity amplitude is slower than if the discontinuity were not considered. This slower falloff is seen in the cutoff region of numerical results, suggesting that the discontinuous method yields a more reasonable choice for $k$ past the peak \cite{steele_lagrange,steele_rootfinding,viergever_Book,deBoer_Rootfinding}. Comparison to numerics is present in the following subsection.

\par{{Recall that these algorithms are designed to solve a root-finding problem for function $f$ of Eqn \ref{rootfind}. The root loci for $f$} in the upper-right panel show that for the discontinuous method, the traversal of the locus is halted as the root pattern begins to appear sparser (i.e. faster change in $k$). As described in Sec \ref{sec:implement}, the discontinuous algorithm then assumes a small negative imaginary root (transition shown by the dashed blue arrow), which yields less rapid falloff in the cutoff region than the larger negative imaginary component found by following the locus continuously.}

\par{The bottom-right panel serves to show that the two methods are both correctly converging to roots of $f$ at each given $x$. Using both methods, the value of $f(kh)$ is less than $10^{-10}$ in magnitude at all $x$ -- this stresses the fact that not all roots lie on the same continuous locus.}

\par
\begin{figure}[ht!]
    \centering
    \includegraphics[width = .7\textwidth]{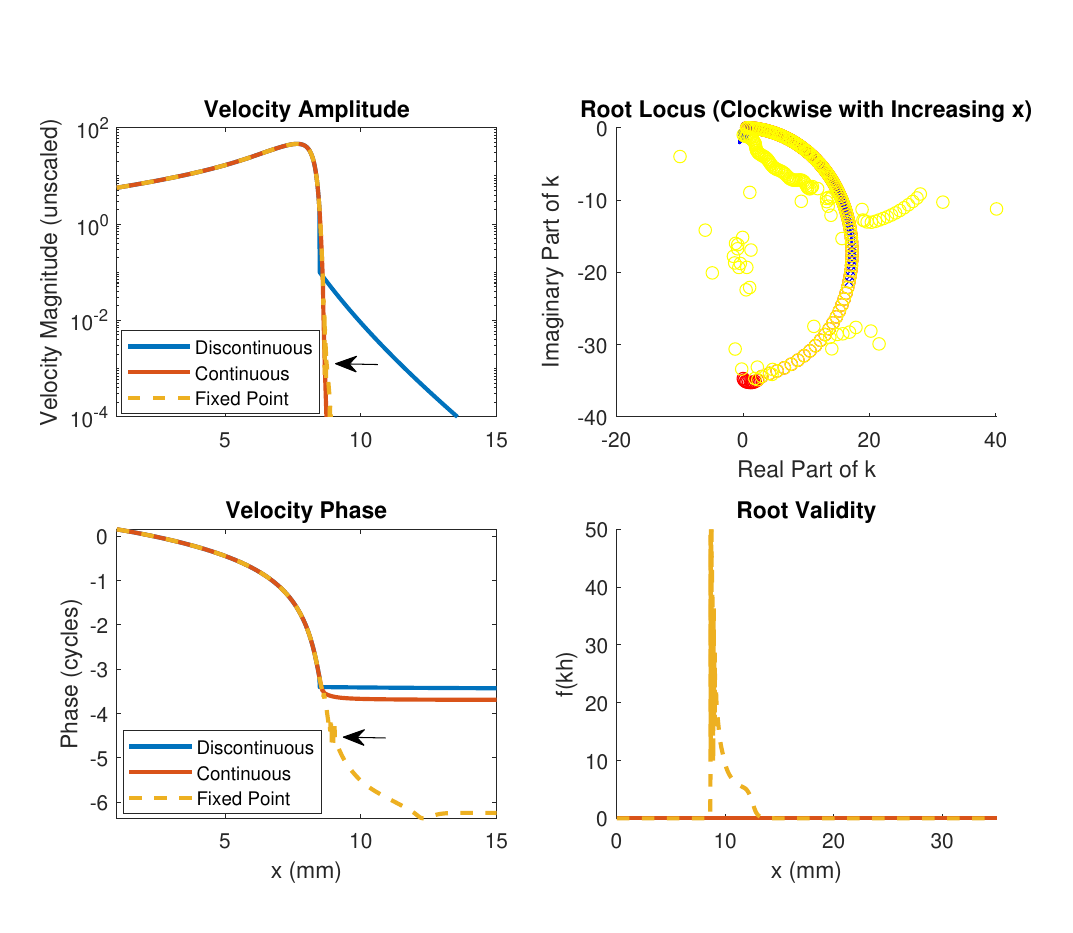}
    \caption{Same as Fig \ref{fig:disc}, but including also the results of the fixed point algorithm as yellow curves in each subplot. Points to note about the fixed point algorithm include the erratic behavior of the root loci and the failure to converge to a proper root near the transition between the peak and cutoff regions. Black arrows in the velocity response plots indicate positions at which this erratic behavior occurs.}
    \label{fig:sp}
\end{figure}

{Fig \ref{fig:sp} shows these same results, but now alongside the results obtained via the fixed point method (Alg \ref{alg::stable}). These results show similar behavior in velocity magnitude to the continuous $x$-stepping solution, but the phase accumulates more cycles.}

\par{Observation of the root locus and $f(kh)$ for the fixed point algorithm reveals strange behavior in the cutoff region. While the fixed point method's root locus follows that of the $x$-stepping method for a large range of $x$, it erratically jumps around the complex plane (including to the third quadrant) past the peak. This corresponds to a non-zero value of $f(kh)$ at these positions as well (see the bottom-right panel), showing that the algorithm has not correctly converged to a root of the function.}

\par{This is anecdotal justification of the validity of this method in the long-wave and short-wave regions, but not in the cutoff region -- a drawback of the fixed point method. The failure of the fixed point algorithm to converge in the cutoff region has been noted before, e.g. in App D of \cite{earhorn}. Still, due to the relative speed of this method's convergence, and its not needing a fine resolution for $x$, it has seen use in many modern works where performance in the cutoff reason is not critical to model results \cite{earhorn,altoe_2022,Shera_Altoe_2023}.}

\subsection{WKB Solutions for the 2-D Box Model}

In the previous subsection, it was shown that the the wavenumber-finding algorithm that accounts for discontinuities in $k$ both converges to roots across space and yields velocity responses that qualitatively resemble numerical results past the peak. This informs the choice to use this algorithm for comparison to numerical results.

\begin{figure}
    \centering
    \includegraphics[width=0.6\textwidth]{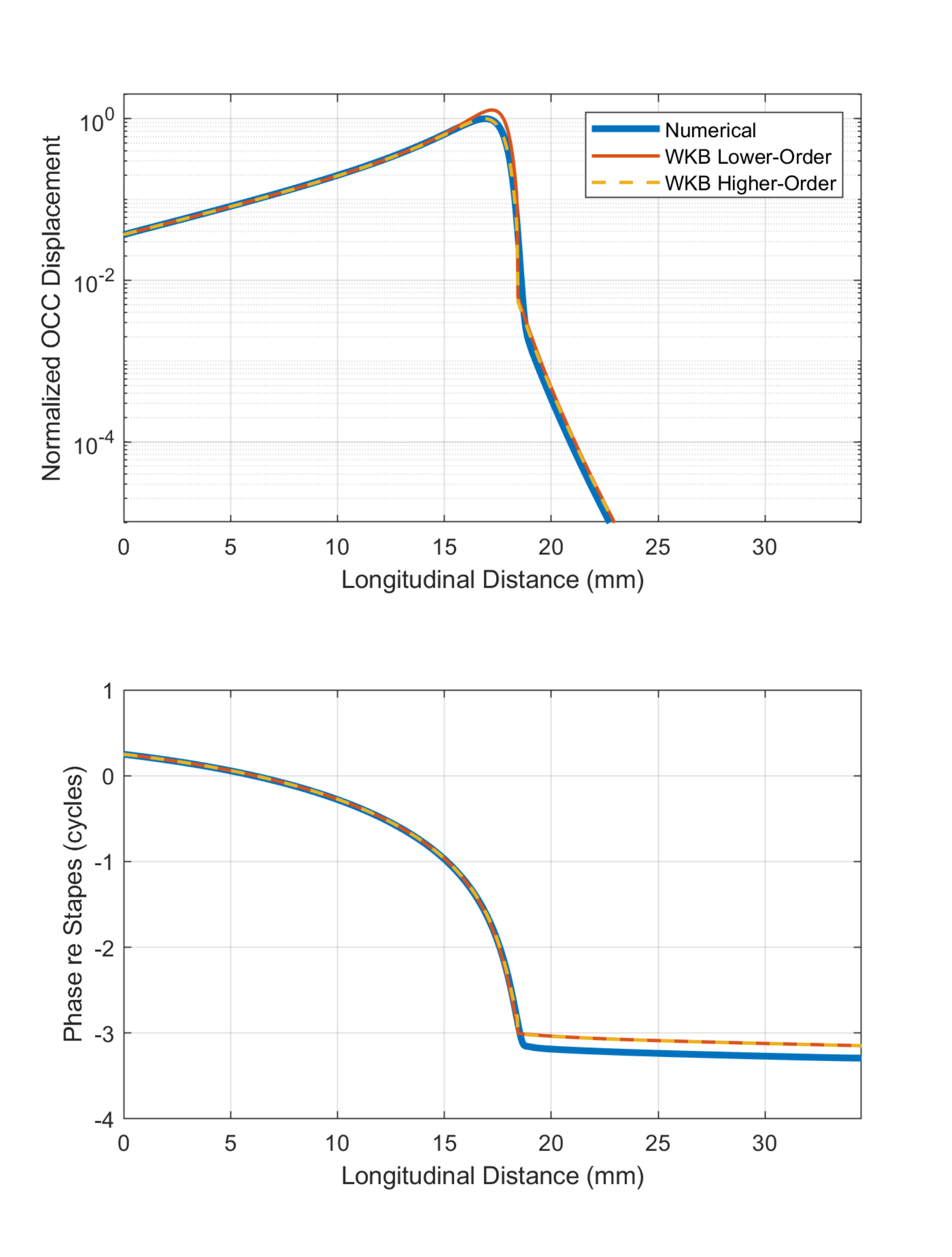}
    \caption{Comparison of numerical solutions to the 2-D box model with lower-order (Eqn \ref{WKB2Dpressure}) and higher-order (Eqn \ref{duifhuisKING}) WKB approximate solutions in response to a 2 kHz stimulus.}
    \label{fig:2d_num}
\end{figure}

In Sec \ref{sec:2d}, two WKB approximate solutions were presented -- the lower-order solution of Eqn \ref{WKB2Dpressure} and the higher-order solution of Eqn \ref{duifhuisKING}. Fig \ref{fig:2d_num} shows solutions according to both of these equations alongside numerical solutions to the 2-D BVP.

Both approximate solutions resemble the numerical solutions across space, including at positions past the peak where the slope of the response rapidly changes. This contrasts with the approximate solutions derived from the two alternate root-finding methods, as seen in Fig \ref{fig:sp}. 

The lower-order solution slightly overestimates tuning at the peak as compared to the higher-order solution, but the solutions are otherwise nearly identical. They differ most significantly from numerical solutions in their phase responses. While they are characteristically similar, both approximate solutions lead the numerical solutions by about 0.1 cycles at apical positions where the phase varies slowly.

It is important to note that these results are sensitive to the choice of threshold $T$ in Alg \ref{alg::with_discon}. A large threshold leads to a registration of a discontinuity at a more apical position. This means that the slope of both the magnitude and phase will change at a more apical location than in the numerical solution. Similarly, a lower value of $T$ will move the discontinuity further basal. For reasonable values of $T$, WKB solutions will be qualitatively similar to numerical solutions in magnitude and phase slope, but they may differ quantitatively due to the shift in the position at which the discontinuity is registered.

\pagebreak
\clearpage
\section{Conclusions}
\label{sec:conclusion}

The WKB approximation provides compact closed-form approximate solutions for 1-D and 2-D cochlear macromechanics models that match numerical solutions well within the entire region of response (all $x$ and $\omega$) save a small region near the resonant position/frequency. These solutions are easily implemented and interpreted, and allow qualitative and quantitative insight into the manner by which physical parameters (impedance variations, scala area, etc.) alter the apical-traveling wave. Through the method of WKB wave-space projection, the solutions also facilitate further interpretation of modeled responses as a superposition of apical- and basal-traveling waves, allowing for the study of intracochlear reflections and otoacoustic emissions.

The forms of the WKB approximate solutions for cochlear box models were developed decades ago, so one may reasonably ask: what is the importance of WKB approximate solutions in contemporary times? In fact, many important insights have been gleamed from WKB solutions in recent literature. Below are just a few such contributions from the past three years.

The inclusion of spatially-varying scala dimensions to a 2-D box model as in Eqn \ref{WKB2Dpressure} has been shown by Alto\`e and Shera to be important for the achievement of substantial OCC velocity at the apex in response to stimuli at the base \cite{earhorn}. They analyzed how tapering scala height could introduce an amplification factor that boosts responses at the apex relative to a uniform-height model, resolving losses that occur in the traveling wave as it makes its way to the apex.

Recent micromechanical findings made through optical coherence tomography have also inspired applications of the WKB solutions to the ostensibly macromechanical box model. In particular, motion at the outer hair cell-Deiters cell junction in the organ of Corti appears to move about 90$^\circ$ out of phase with basilar membrane motion within the same longitudinal cross-section. Implementing this as a modification to the impedance term, Alto\`e and Shera have used the WKB solutions as derived in this tutorial to model the impact of such a phenomenon \cite{altoe_2022}. They arrive at an alternative interpretation of cochlear amplification, in which power may be supplied to the fluid rather than directly to the basilar membrane.

Recent work by Sisto \textit{et al.}  used the WKB approximation in studying the level-dependence of the OCC admittance, assumed to arise from outer hair cell motility \cite{sisto_2021,sisto_2023}. Paying special attention to a) the pressure focusing phenomenon described above, and b) the viscosity at the OCC-fluid interface, they have found that substantially level-dependent admittance is not required to obtain the impressive dynamic range of the cochlea.

With much still to learn about the mechanics of the cochlea, the powerful analytic tool offered by the WKB approximation is among the strongest we have due to its interpretability, versatility and simplicity of computation. With the foundations discussed in this tutorial, derivations and implementation details can be modified to tackle contemporary questions as they continue to arise. 

\section*{Acknowledgments}
I would like to thank Dr. Elizabeth S. Olson and the JASA reviewers for providing edits and comments for this tutorial.

\section*{Author Declarations}
The author declares no conflicts of interest. No data was used in this tutorial article. Scripts used in generating the presented figures are available upon request.
\vfill
\pagebreak
\clearpage
\appendix
\renewcommand{\theequation}{A.\arabic{equation}}
\setcounter{equation}{0}
\section{The Higher-Order 2-D Model -- A Series Solution Approach}
\label{app:series}

In this appendix, I provide a derivation of the higher-order ``WKB solution" of Eqn \ref{duifhuisKING} from the main text. The following approach is modified from Viergever's in his 1980 book \textit{Mechanics of the Inner Ear: A Mathematical Approach} \cite{viergever_Book}. It relies on a transformation of the coordinates of the pressure BVP, and subsequent application of a WKB-adjacent ansatz (but not precisely the WKB method).

The method consists of the following steps:
\begin{enumerate}
    \item{Change the variables of the BVP in pressure so that terms relating to the model parameters appear in the PDE rather than only in the boundary conditions.}
    \item{Write a form for the solution to this new PDE as
    \begin{equation*}
        A(x,\zeta)\cosh{[\kappa(x)(H-\zeta)]}e^{jKg(x)}.
    \end{equation*}
    That is, assume that the $z$ (here reparameterized as $\zeta$) contribution is hyperbolic and that there is a wave in $x$. The product with arbitrary $A(x,y)$ means this is done without loss of generality.}
    \item{Assume a series solution for $A$ and plug into the ODE to obtain a system of PDEs.}
    \item{Solve for $A$ up to first order, plug back in to the ansatz and undo the change of variables to solve for pressure.}
\end{enumerate}

A detailed outline is presented below, but certain steps feature highly nontrivial computations. For full exposition of these computations, see  https://github.com/brian-lance/wkb-derivations.

\subsection{Setting up the BVP}

The method followed in this section relies on multiple changes of variables and definitions of new parameters. As such, it can be difficult to keep straight the meanings and units of the various variables and parameters at play. Table \ref{tab::vier} serves as a reference for the objects introduced in the derivation.

\begin{table}
    \centering
    \begin{tabular}{|c||c|c|} \hline 
         Symbol & Significance & Units\\ \hline \hline
         $Z_0$ & Arbitrary reference impedance used to the simplify series solution. & Pa$\cdot$s/mm \\ \hline
         $K$ & Reference wavenumber used to simplify the series solution.  & 1/mm \\ \hline
         $f^2(x)$ & $Z_0/Z_{OC}(x)$, used so simplify $x$-dependence of the PDE. & Unitless \\ \hline
         $\zeta$ & $Kz$, Nondimensionalized transverse coordinate. & Unitless \\ \hline
         $H$ & $Kh$, Nondimensionalized scala height. & Unitless \\ \hline
         $Q(x,\zeta)$ &  Pressure written in terms of the nondimensionalized transverse coordinate. & Pa \\ \hline
         $A(x,\zeta)$ & Auxiliary pressure variable that controls the magnitude of pressure at the OCC,& \\ & to be solved for in the simplified BVP. & Pa \\ \hline
         $\kappa(x)$ & Controls the $x$-dependence of transverse pressure variations, & \\ & to be solved for in the simplified BVP. & Unitless\\ \hline
         $g(x)$ & Controls the wavenumber of the traveling wave, & \\ & to be solved for in the simplified BVP. & mm \\ \hline
    \end{tabular}
    \caption{Symbols introduced in the derivation of the model equations in the series solution approach, along with their significance and units.}
    \label{tab::vier}
\end{table}

Viergever begins with the 2-D box model Laplace equation BVP in $P(x,z)$, then performs a change of variables. To start, we define a reference impedance $Z_0$ which is some arbitrary constant. We define also $f^2(x) = Z_0/Z_{OC}(x)$ and a reference wavenumber $K^2 = -2j\omega\rho/hZ_0$. Recalling that $P = -\rho \dot{\phi}$,  the boundary condition at the OCC is
\begin{equation}
    \frac{\partial P}{\partial z} + hK^2f^2(x)P=0,   z=0.
\end{equation}
Note that $K$ is not a function of $x$.

Further reparameterizing the $z$ coordinate and defining a reparameterized pressure, $Q$, as 
\begin{equation}
    \zeta = Kz,  H=Kh,  Q(x,\zeta) = P(x,z),
\end{equation}
the BVP in terms of $Q$ is
\begin{align}
    \frac{\partial^2 Q}{\partial x^2} + K^2\frac{\partial^2 Q}{\partial \zeta^2} = 0,\\
    \frac{\partial Q}{\partial \zeta}\bigg|_{\zeta=H} = 0,\\
    \frac{\partial Q}{\partial \zeta}\bigg|_{\zeta=0} + Hf^2(x)Q(x,0) = 0.
\end{align}

\par{The solution to the above PDE is artificially represented in a form resembling what the solutions are expected to be by intuition about the solutions of the Laplace equation. In particular, $Q$ is written as
\begin{equation}
    Q(x,\zeta) = A(x,\zeta;K)e^{jKg(x)}\cosh{[\kappa(x)(H-\zeta)].}
    \label{waveishguess}
\end{equation}

 The exponential suggests a traveling wave in $x$, where $A$ modulates the amplitude of this wave. However, this is not actually an assumption of a wave solution -- as $A$, $g$ and $\kappa$ are unknown functions of $x$ (and $\zeta$, for $A$), any function can be represented in this fashion without loss of generality.}

\par{One might wonder why we have chosen to introduce so many new terms into these equations. While this may initially seem to complicate the BVP, it eventually leads to the most mathematically tractable solution method.}
\par{Alongside Table \ref{tab::vier}, it may help to ``look into the future" to see what these newly defined variables will become. The variable $\kappa$ will be found to be the nondimensionalized wavenumber and $g$ will be found to be the integral of the wavenumber. The free parameter $K$ will eventually be the variable of our formal power series (similar to $\delta$ from Eqn \ref{WKB_withdelta}).}
 
Plugging this form of $Q$ into the BVP, we can find an equivalent BVP in terms of $A$. Writing $a(x,\zeta) = \kappa(x)(H-\zeta)$ to simplify notation, we arrive at the following PDE and boundary conditions:
\begin{align}
\begin{split}
    K^2&\bigg[(\kappa^2 - g'^2)A\cosh{a} + \frac{\partial^2 A}{\partial \zeta^2}\cosh{a} -2\kappa \frac{\partial A}{\partial \zeta}\sinh{a} \bigg]+\\
    +jK&\bigg[g''A\cosh{a} + 2g'\frac{\partial A\cosh{a}}{\partial x} \bigg] + \frac{\partial^2 A\cosh{a}}{\partial x^2}=0,
    \end{split}
    \label{PDEinA}
\end{align}
\begin{equation}
    \frac{\partial A}{\partial \zeta}\bigg|_{\zeta = 0} -\kappa A(x,0)\tanh{a(x,0)} + Hf^2A(x,0) = 0,
\end{equation}
\begin{equation}
    \frac{\partial A}{\partial \zeta}\bigg|_{\zeta = H} = 0.
\end{equation}

Solving this ODE in the auxiliary pressure $A$ is the new goal. With a solution for $A$, we can find $Q$ and finally $P$.

\subsection{A Series Solution for Auxiliary Pressure}
A formal power series solution in the form
\begin{equation}
    A(x,\zeta;K) = A_0(x) + \sum_{n=1}^\infty \frac{1}{(jK)^n}A_n(x,\zeta)
\end{equation}
is assumed, with monotonic decrease in magnitude of terms and their derivatives in increasing $n$, and allowing for termwise differentiation. This form of the solution is not quite the WKB ansatz, but the logarithm of such a solution with $\delta = jK$. This is motivated by Keller's approach to surface waves \cite{keller}.

This ansatz is plugged into the PDE for $A$ in Eqn \ref{PDEinA}, resulting in a system of infinitely many PDEs of which we consider only the PDEs including $A_0$ and $A_1$ (justified by the terms and their derivatives being assumed to decrease monotonically). The resulting system of differential equations is
\begin{equation}
    g'^2(x) = \kappa^2(x),
    \label{recurse0}
\end{equation}

\begin{equation}
    \cosh{a}\frac{\partial^2 A_1}{\partial \zeta^2} - 2\kappa\sinh{a}\frac{\partial A_1}{\partial \zeta} =g''A_0\cosh{a} + 2g'\frac{\partial A_0 \cosh{a}}{\partial x}.
    \label{seriespde0}
\end{equation}

Application of boundary conditions gives
\begin{equation}
    \frac{\partial A_1}{\partial \zeta}\bigg|_{\zeta=H} = 0,
    \label{Hboundary}
\end{equation}
\begin{equation}
    \frac{\partial A_1}{\partial \zeta}\bigg|_{\zeta=0} = 0,
\end{equation}
\begin{equation}
    \kappa\tanh{\kappa H} = Hf^2.
    \label{kapparelation}
\end{equation}
The final equation resembles the dispersion relation derived from the WKB method in Sec \ref{sec:2d}.

Eqn \ref{recurse0} is solved by
\begin{equation}
    g(x) = \pm \int_0^x \kappa(\xi) d\xi + C
    \label{geqn}
\end{equation}
for arbitrary constant $C$. This resembles the characteristic WKB phase term. 

\subsection{Finding a First Approximation for Pressure}

Solving Eqn \ref{seriespde0} is nontrivial, as it contains both $A_0$ and $A_1$. Solution for $A_0$ requires clever substitutions\footnote{This is outlined at https://github.com/brian-lance/wkb-derivations}. I find
\begin{equation}
    A_0 = C(\kappa H + \sinh{\kappa H}\cosh{\kappa H})^{-1/2},
\end{equation}
for arbitrary $C$. 

Theoretically this facilitates solution for $A_n$ for any $n$ as well. On the other hand, the series approximation gives that the higher $n$ terms should be small if $K$ is large relative to its own rate of change (analogous to the WKB assumption).

Ignoring $A_n$ for $n\geq 1$ gives a first approximation for $Q$ by putting $A\approx A_0$. Using Eqn \ref{geqn} for $g$, there are two possible solutions: 
\begin{equation}
    Q_{\pm}(x,\zeta) = C_{\pm}(\kappa H + \sinh{\kappa H}\cosh{\kappa H})^{-1/2} e^{\pm j K \int_{0}^x\kappa(\xi) d\xi}\cosh{[\kappa(x)(H-\zeta)]} + O(1/K).
\end{equation}

The reference constant $K$ was defined as $K^2 = -2j\rho\omega/hZ_0$, where $Z_0$ was a \textit{second} reference constant so that $f^2 = Z_0Y_{OC}$. Because $Z_0$ was entirely arbitrary, I am free to choose $Z_0 = -2j\rho\omega h^{-1}$ so that $K=1$ mm$^{-1}$. This also gives $H=1$ mm$^{-1} \times h$ [unitless], $\zeta=1$ mm$^{-1} \times z$ [unitless] and $Q = P$ [Pa].

I define $k = 1$ mm$^{-1} \times \kappa$. In this light, the $\kappa$ relation from Eqn \ref{kapparelation} becomes
\begin{equation}
    k \tanh{k h} = -2j\rho \omega Y_{OC}.
    \label{dispViergever}
\end{equation}
This is \textit{precisely} the dispersion relation derived through the WKB method (Eqn \ref{dispersion}), where $k$ is the wavenumber with units mm$^{-1}$.

The first approximation for pressure with arbitrary constants $C_{\pm}$ is now
\begin{equation}
    P(x,z) = (k h + \sinh{k h}\cosh{k h})^{-1/2}\cosh{[k(x)(h-z)]}\bigg[ C_+ e^{j \int_{0}^x k(\xi) d\xi} + C_- e^{- j \int_{0}^x k(\xi) d\xi}\bigg].
\end{equation}

To find the constants, the two $x$-boundary conditions are used:
\begin{equation}
    \frac{1}{h}\int_0^h P(0,z) dz = P_{OW},    \frac{\partial P}{\partial x}\bigg|_{x=L} = 0,
\end{equation}
where $L$ is the length of the cochlea and $P_{OW}$ is the average pressure at the stapes. 

After some computation\footnote{These can be found at https://github.com/brian-lance/wkb-derivations}, assuming that the backwards traveling wave is negligible, we achieve
\begin{equation}
     P(x,z) = \frac{P_{OW} k_0 h}{\cosh{k(x) h} \tanh{k_0 h}}\sqrt{\frac{k_0 h\text{sech}^2 k_0 h + \tanh{k_0 h}}{k(x) h \text{sech}^2 k_(x) h + \tanh{k(x) h}}} \cosh{[k(x)(h-z)]} e^{-j\int_0^x k(\xi) d\xi}.
\end{equation}
This is precisely Eqn \ref{duifhuisKING}.

\pagebreak
\clearpage
\printbibliography

\end{document}